\documentclass[a4paper,11pt]{article}
\pdfoutput=1 

\usepackage{jinstpub} 
\usepackage{placeins}
\usepackage{xcolor}
\usepackage{amsfonts}
\usepackage{booktabs}
\usepackage{siunitx}
\usepackage{tikz}
\newcommand{\ras}[1]	 {\renewcommand{\arraystretch}{#1}} 

\title{\boldmath Maximum performance of strange-jet tagging at hadron colliders}

\author{Johannes Erdmann, Olaf Nackenhorst and Sonja Verena Zei{\ss}ner\footnote{Corresponding author}}
 \affiliation{Lehrstuhl f\"ur Experimentelle Physik IV, TU Dortmund, Otto-Hahn-Stra{\ss}e 4a, 44227 Dortmund, Germany}

\emailAdd{sonja.zeissner@tu-dortmund.de}

\abstract{
  The maximum achievable performance of strange-jet tagging at hadron colliders and the loss in performance in different detector designs is estimated based on simulated truth jets from strange-quark and down-quark hadronisation. Both jet types are classified with a recurrent neural network using long short-term memory units, at first using all available truth particles and then applying selections to study the impacts of ideal tracking detectors, Cherenkov detectors, and calorimeters. Additionally, a manual reconstruction of strange hadron decays such as $K_S\rightarrow \pi^+ \pi^-$ from charged tracks is considered.}


\begin{document}
\maketitle
\flushbottom

\section{Introduction}
\label{sec:intro}

The identification of the flavour of a jet, i.e.~the type of elementary particle from which the jet originated, is a crucial experimental technique at collider experiments, such as experiments at the Large Hadron Collider (LHC)~\cite{Evans2008}. 
The identification of jets that originate from $b$-quarks~\cite{Abreu:1999qh} is an established technique at the ATLAS~\cite{Aad:2019aic} and CMS~\cite{Sirunyan:2017ezt} experiments with a multitude of applications. Methods for the identification of jets from $c$-quarks~\cite{Sirunyan:2017ezt,Aaboud:2018fhh} and gluons~\cite{Pumplin:1991kc,CMS:2013kfa,Aad:2014gea,Komiske:2016rsd,Cheng:2017rdo,Metodiev:2017vrx,Dery:2017fap,Metodiev:2018ftz,Komiske:2018vkc,Komiske:2018cqr,Kasieczka:2018lwf,Mikuni:2020wpr,Kasieczka:2020nyd} have also been developed in recent years. 
If a particle carries a relatively high momentum w.r.t.~its mass, the decay products of this particle are reconstructed as a single large-radius jet. The substructure of these jets is for example used for the identification of hadronic decays of top quarks~\cite{Aad:2016pux,Kasieczka:2017nvn,Butter:2017cot,Aaboud:2018psm,Macaluso:2018tck,Dillon:2019cqt,Diefenbacher:2019ezd,Sirunyan:2020lcu,Chakraborty:2020yfc}, $W$ and $Z$ bosons~\cite{Khachatryan:2014vla,Cogan:2014oua,Aad:2015rpa,Aad:2015eax,Datta:2017rhs,Aaboud:2018psm,Chen:2019uar,Sirunyan:2020lcu,Ju:2020tbo}, and Higgs bosons~\cite{Lin:2018cin,Aad:2019uoz,Chakraborty:2019imr,Moreno:2019neq,Sirunyan:2020lcu,Ju:2020tbo}.

Algorithms for the identification of jets from strange-quarks ($s$-jets) are currently not in use at the LHC. Such a strange-tagging (or $s$-tagging) algorithm, however, would complement the existing algorithms for the identification of the jet flavour and could open new opportunities in the analysis of the data, such as searches for the decays $t\rightarrow W^+ s$~\cite{Ali:2010xx} and $H\rightarrow s\bar{s}$~\cite{Duarte-Campderros:2018ouv}, especially for large data sets to be collected in future high-luminosity LHC runs.
Currently, $s$-jets can be distinguished from jets that stem from $b$- and $c$-quarks making use of $b$- and $c$-tagging algorithms. Since these algorithms treat $s$-jets and jets from other light partons as a collective background, the separation of these two types is the main challenge of an $s$-tagging algorithm.  At the Stanford Linear Collider and the Large Electron-Positron Collider~\cite{Myers:1990sk}, Cherenkov detectors were used to identify $s$-jets~\cite{Kalelkar:2000ig,Abreu:1999cj}, exploiting their identification power for charged kaons. More recently, machine-learning methods have been explored for $s$-tagging based on tracking~\cite{Erdmann:2019blf} and additional calorimeter information~\cite{Nakai:2020kuu}. However, these studies show that it is challenging to achieve a good separation of $s$- and other light jets --- in particular compared to the excellent identification that is achieved for $b$- and $c$-jets. One reason is that strange-hadrons, such as kaons and $\Lambda$ baryons, do not only appear in the fragmentation of $s$-quarks but also are frequently part of $d$- and $u$-jets. Another reason is that the identification of these strange-hadrons is experimentally challenging: (1)~$K_S$ mesons and $\Lambda$ baryons often decay within the inner tracking detectors of collider experiments, but their identification is only possible if they decay to two charged particles and the tracks of both decay products are well-measured so that their invariant mass can be reconstructed; (2)~$K_L$ and $K^\pm$ mesons have long lifetimes and mostly do not decay before producing hadronic showers in the calorimeters, which are then difficult to distinguish from the showers of other hadrons; (3)~tracks from $K^\pm$ mesons could, in principle, be distinguished from other charged particles with the use of Cherenkov detectors or $\text{d}E/\text{d}x$ measurements. Cherenkov detectors are currently not used nor foreseen for the general-purpose detectors at the LHC. In addition, particle identification based on $\text{d}E/\text{d}x$ measurements is only feasible for heavy particles for the large momenta common at the LHC.

The goal of this paper is to determine the best achievable $s$-tagging performance in an ideal environment and study the impact of fragmentation and other experimental effects on this performance. Assuming an ideal detector that can perfectly measure both the four-momentum as well as the trajectory of both charged and neutral particles, the maximum performance of an $s$-tagger is solely based on the differences in the fragmentations of $s$- and $d$-jets. While this clearly cannot be achieved in a realistic experiment, it defines a useful benchmark and an upper bound for more realistic $s$-tagging scenarios. In order to determine this benchmark, neural networks are trained based on Monte-Carlo simulations in order to separate $s$- and $d$-jets using the properties of their constituent particles\footnote{The $s$-tagging performance for a tagger based on tracking information was found to be similar for $d$- and $u$-jets~\cite{Erdmann:2019blf}, so that we limit ourselves in this study to $d$-jets as the only background class.}. 
The application of machine learning in high-energy physics has proven to be a successful approach, especially for the classification of jets \cite{Cogan:2014oua,deOliveira:2015xxd,Guest:2016iqz,Komiske:2016rsd,Aguilar-Saavedra:2017rzt,Butter:2017cot,Cheng:2017rdo,Datta:2017rhs,Dery:2017fap,Kasieczka:2017nvn,Louppe:2017ipp,Metodiev:2017vrx,Sirunyan:2017ezt,Aaboud:2018fhh,Aaboud:2018psm,Heimel:2018mkt,Kasieczka:2018lwf,Komiske:2018cqr,Komiske:2018oaa,Komiske:2018vkc,Lin:2018cin,Macaluso:2018tck,Metodiev:2018ftz,Aad:2019aic,Belayneh:2019vyx,Chakraborty:2019imr,Chen:2019uar,Diefenbacher:2019ezd,Dillon:2019cqt,Erdmann:2019blf,Moreno:2019bmu,Moreno:2019neq,Qu:2019gqs,Bols:2020bkb,Chakraborty:2020yfc,Ju:2020tbo,Kasieczka:2020nyd,Mikuni:2020wpr,Nakai:2020kuu,Sirunyan:2020lcu}. For the following studies, long short-term memory recurrent neural networks (LSTM)~\cite{Hochreiter:1997yld} are used, because they provide a sufficiently complex and flexible model, which is capable of capturing differences in the properties of a variable number of particles constituting the jet. The impact of various experimental effects is then studied by considering several scenarios with different combinations of idealized detector components. These provide an estimate of the maximum performance that can be achieved for $s$-tagging based on tracking information, additional Cherenkov detectors, and with calorimeter information. These studies provide guidance for further developments for $s$-jet tagging based on these detector components.

\section{Simulated samples}

To compare jets originating from $s$- and $d$-quarks, di-quark production events in proton--proton collisions, i.e. $pp \rightarrow s\overline{s}$ and $pp \rightarrow d\overline{d}$, are simulated at $\sqrt{s} = 13\text{\,TeV}$. The matrix element is simulated at leading order in $\alpha_S$ with MadGraph\_aMC@NLO version 2.6.7~\cite{Alwall:2011uj} using the NNPDF2.3LO set of parton distribution functions~\cite{Ball:2012cx}. The parton shower and hadronisation is simulated with Pythia~8.2.35~\cite{Sjostrand:2014zea}. Since only one hadronisation model is used, future studies will be necessary to evaluate the impact of using different hadronisation models. The events are not passed through any detector simulation, hence it is assumed that all quantities are measured perfectly. All resulting particles with a non-zero lifetime -- independent of their electric charge or other properties -- are clustered into jets with the anti-$k_{\mathrm{T}}$-algorithm~\cite{Cacciari:2008gp} with a radius parameter of $R=0.4$ using FastJet~\cite{Cacciari:2011ma}. In order to label the flavour of the jets, the $s$- and $d$-quarks from the matrix element are ghost-matched~\cite{Cacciari:2008gn} to the resulting jets.

In total, 2.2 million jets of each jet type are produced. Their transverse momentum, $p_{\text{T}}$, and direction is determined by summing the four-vectors of all jet constituents that have their origin in the primary vertex\footnote{For the description of all events, modified spherical coordinates ($r,\phi,\eta$) are used, where $(r,\phi)$ span the plane transverse to the proton beam. The polar angle $\theta$ is expressed in terms of the pseudorapidity $\eta = -\ln \tan\left(\frac{\theta}{2}\right)$. The angular distance between two objects is defined as $\Delta R = \sqrt{(\Delta\eta)^2+(\Delta\phi)^2}$.}.  A lower cut of $p_{\text{T}}>\SI{25}{GeV}$ -- which in its magnitude is typical for LHC experiments -- is applied for all jets, yielding an average $p_{\text{T}}=37\,\text{GeV}$. In order to remove small kinematic differences between $s$- and $d$-jets, the $p_{\text{T}}$ and $\eta$ distributions of $d$-jets are reweighted to match the ones of $s$-jets. These weights are applied to all $d$-jets during the training and evaluation of the neural networks and in all plots.

\begin{figure}[t]
\centering
\includegraphics[width=.6\textwidth]{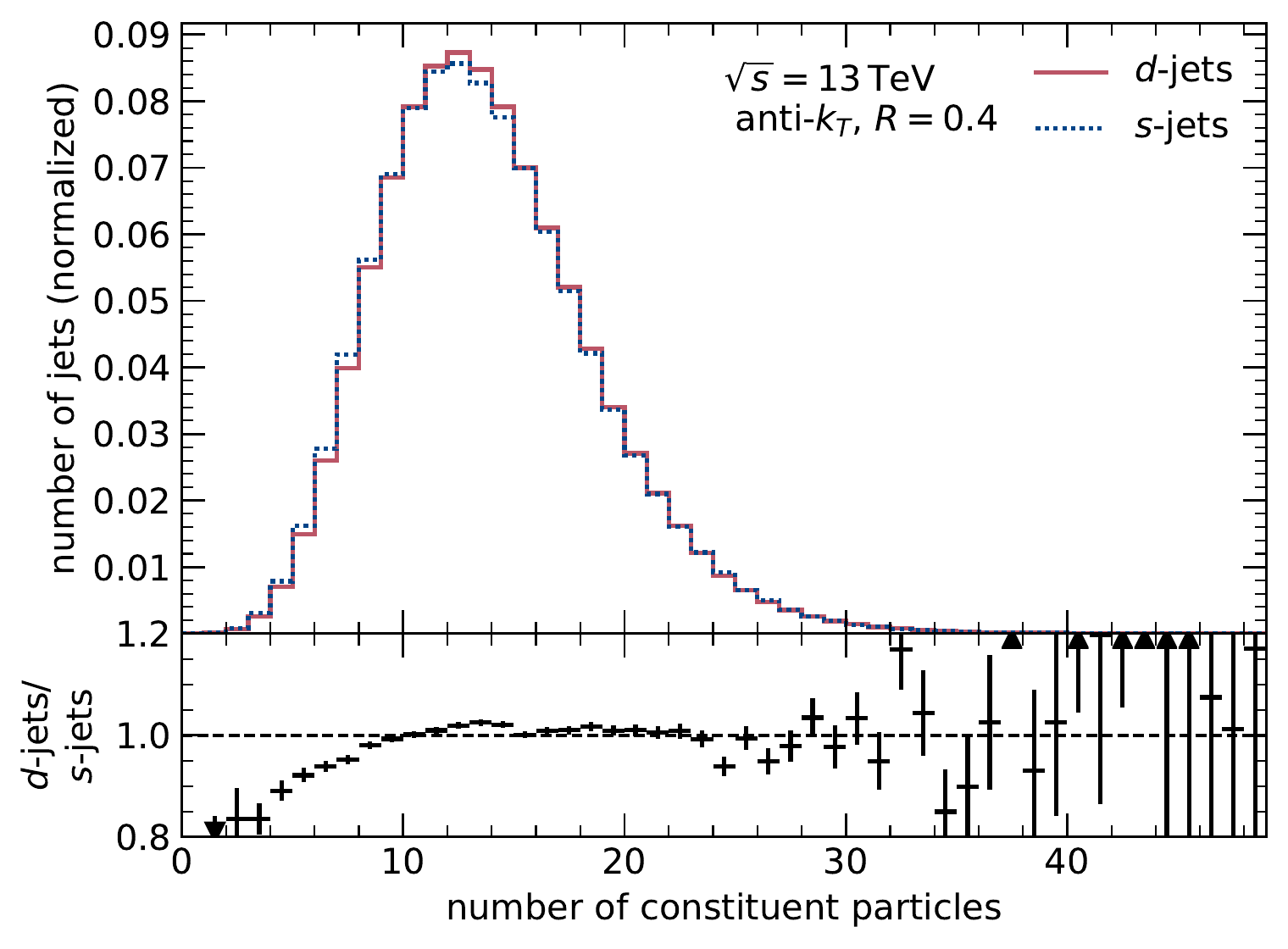} 
\caption{Distribution of the number of constituent particles per $s$- and $d$-jet clustered from all particles with a lifetime greater zero.}
\label{FIG:noparticles}
\end{figure}

\begin{figure}[t]
\centering
\includegraphics[width=.6\textwidth]{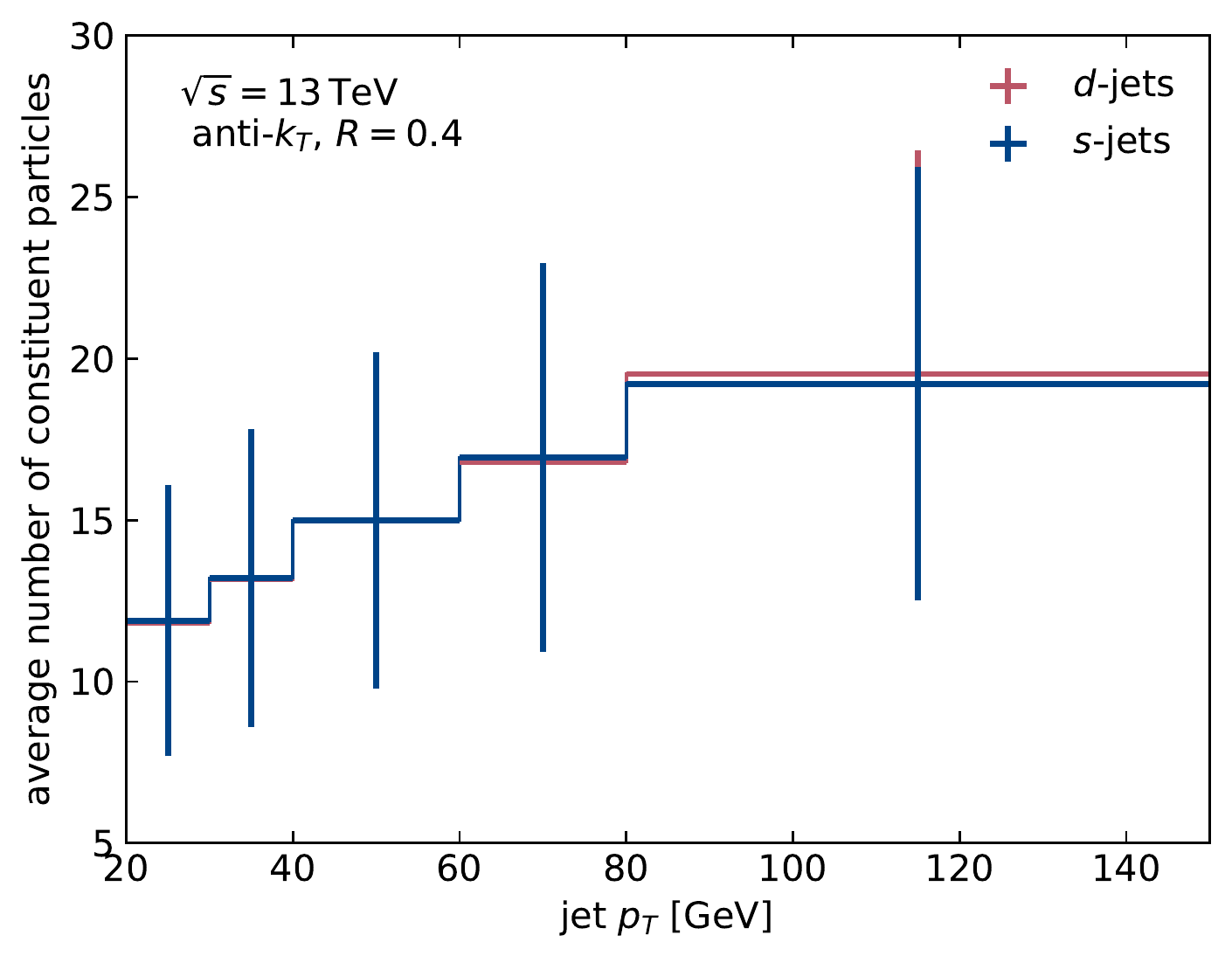} 
\caption{Average number of constituent particles for different transverse jet momenta of $s$- and $d$-jet. The error bars correspond to the standard deviation of the number of constituent particles in each bin (and not the standard deviation of the mean value).}
\label{FIG:noparticles_pt}
\end{figure}

All jet constituents are ordered by energy, when they are given as input to the neural network. An alternative ordering by the particles radius of origin $r_0$ was tested and found to result in a similar $s$-tagging performance. The number of constituent particles per jet across all jet $p_{\mathrm{T}}$ is shown in Figure~\ref{FIG:noparticles}. Figure~\ref{FIG:noparticles_pt} shows the average number of constituent particles for different transverse jet momenta $p_{\mathrm{T}}$. 
The number of constituent particles is similar for $s$- and $d$-jets. It increases for increasing jet $p_{\mathrm{T}}$.  Because only a very small fraction of jets has more than 50 constituent particles, any additional particles are not considered in the following.

\begin{figure}[t]
\centering
\includegraphics[width=.6\textwidth]{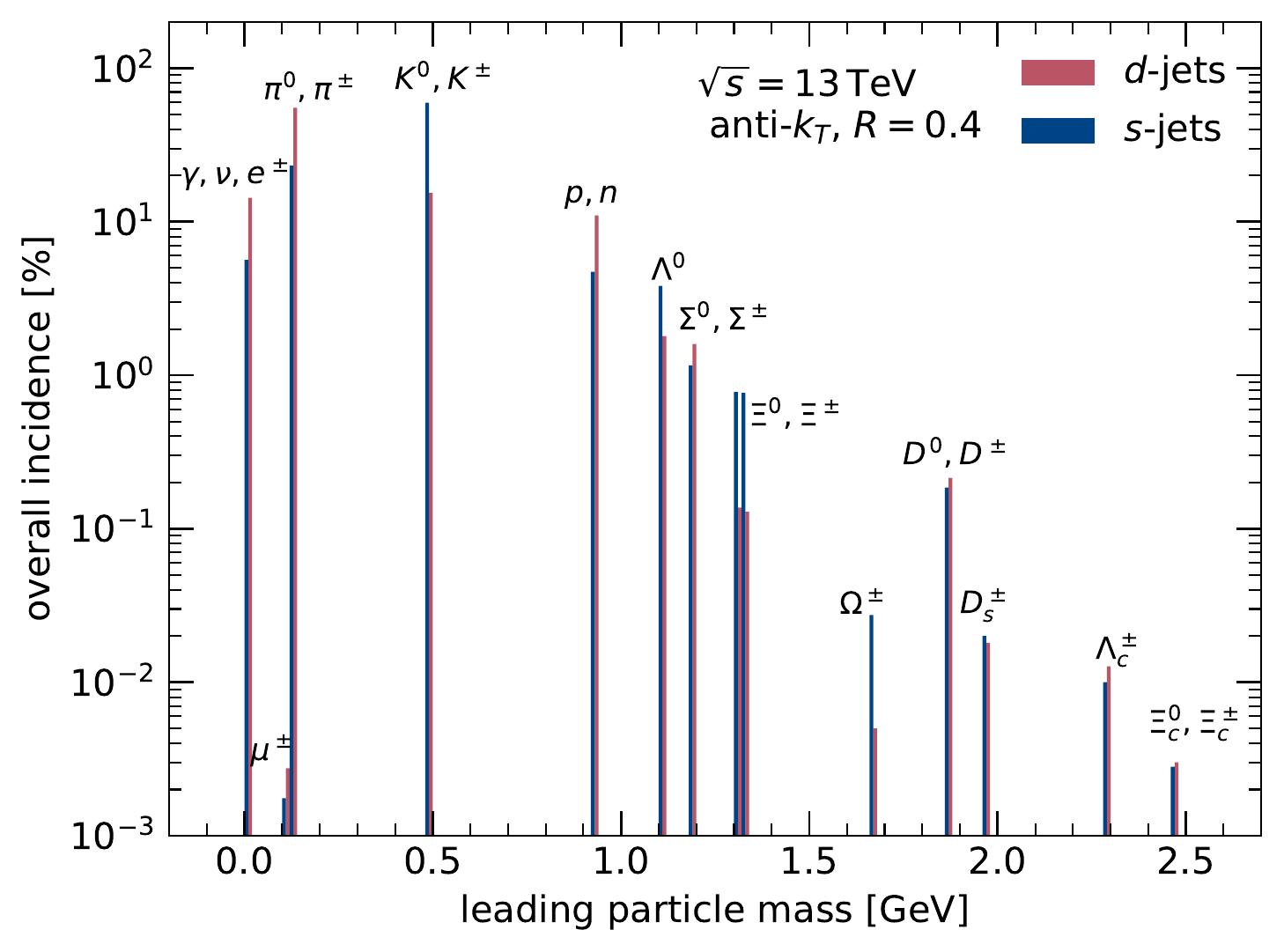} 
\caption{Mass spectrum of the jets' constituent particle leading in energy for $s$- and $d$-jets.}
\label{FIG:particlemassleading}
\end{figure}

\begin{figure}[t]
\centering
\includegraphics[width=.6\textwidth]{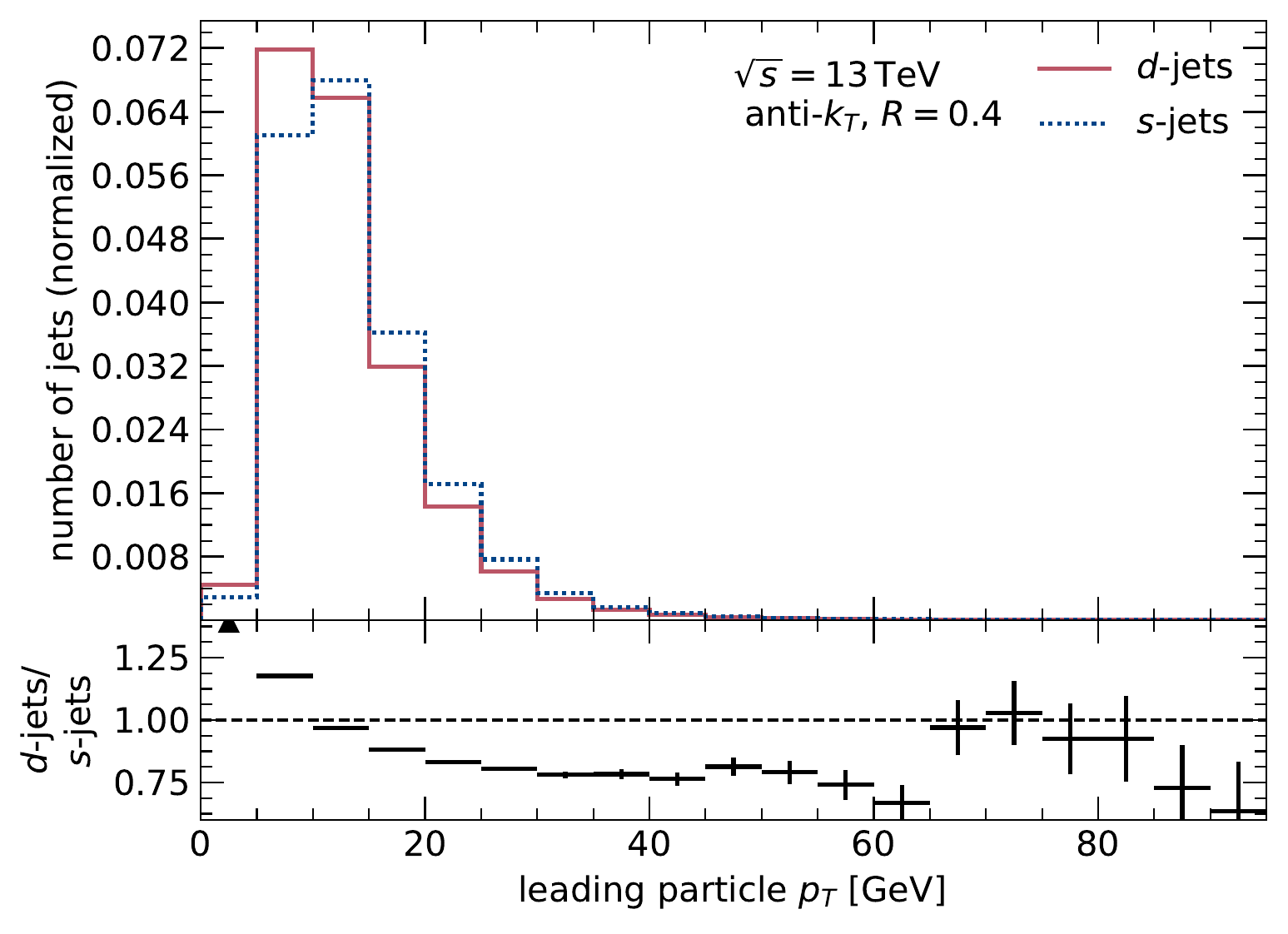} 
\caption{Distribution of the $p_{\mathrm{T}}$ of the jets' constituent particle leading in $p_{\mathrm{T}}$ for $s$- and $d$-jets.}
\label{FIG:particleenergy}
\end{figure}

Figure~\ref{FIG:particlemassleading} shows the mass and the type of the leading constituent particle in the jets. For $s$-jets, the leading particle tends to be a strange-hadron (mostly kaons), while for $d$-jets, the leading particle is more likely to be a light hadron --- in particular a pion. While, on average, the type of the leading particle differs between $s$- and $d$-jets, its kinematic properties are similar. As an example, the transverse momentum of the leading particle is shown in Figure~\ref{FIG:particleenergy}. The leading particles of $s$-jets tend to carry  more momentum than those of $d$-jets. This is in line with the fact that kaons produced in $pp$ collisions tend to have larger transverse momentum as known from fragmentation functions determined at the same center-of-mass energy \cite{Sirunyan:2017zmn}. 

\begin{figure}[t]
\centering
\includegraphics[width=.6\textwidth]{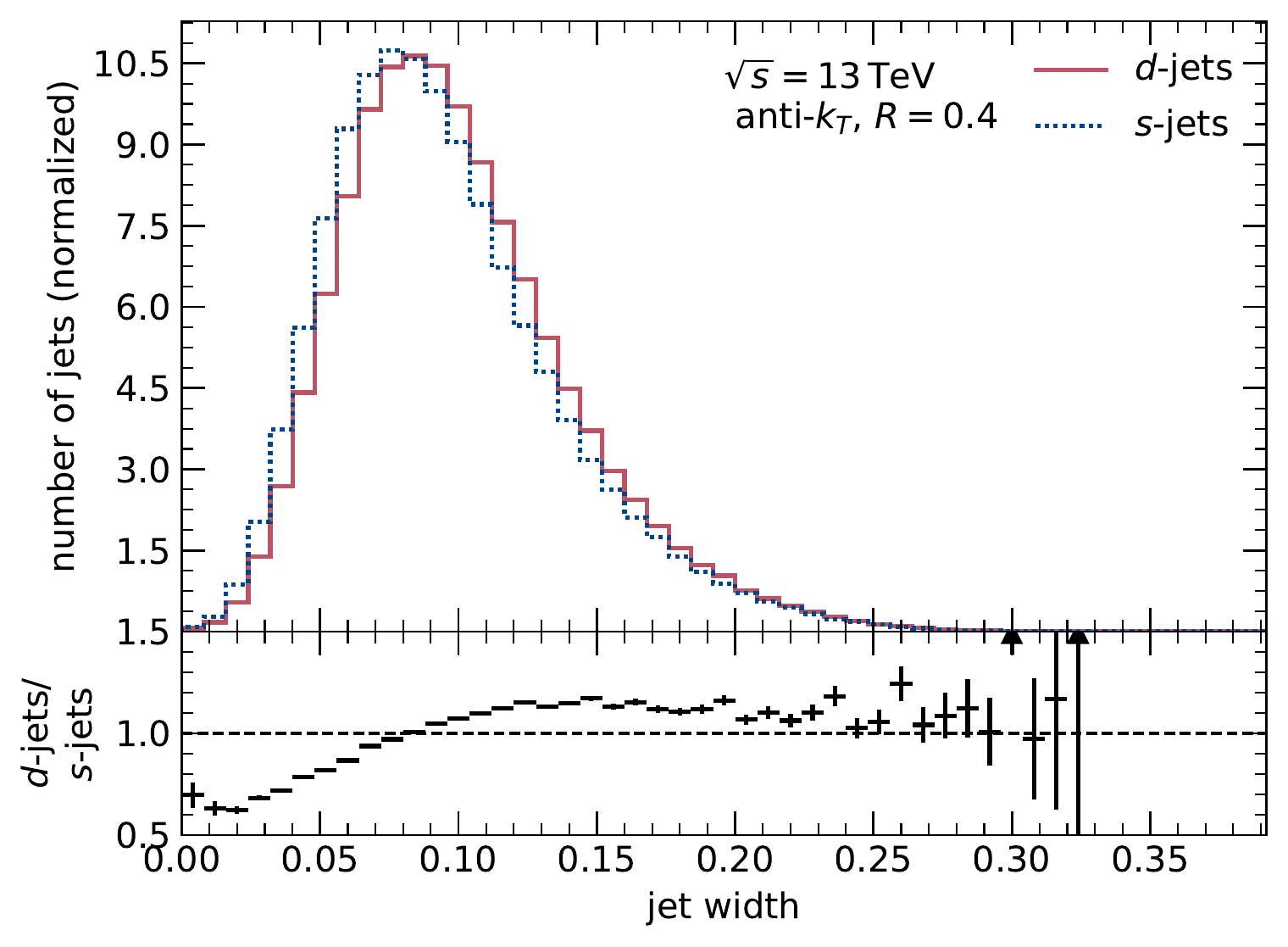} 
\caption{Distribution of the jet width of $s$- and $d$-jets.}
\label{FIG:jetwidth}
\end{figure}

Some differences between $s$- and $d$-jets are, for example, observed in the width of the jets, which is defined as
\begin{equation}
\text{jet width} = \frac{\sum\limits_{i} p_{\text{T}}^i \; \Delta R(i,\text{jet})}{\sum\limits_i p_{\text{T}}^i}\, ,
\end{equation}
where $i$ runs over all particles. The jet width, as shown in Figure~\ref{FIG:jetwidth}, tends to be on average narrower for $s$-jet than for $d$-jets. However, the comparison of the two distributions also demonstrates that even complex and engineered features, like the jet width, have little discrimination power, which makes $s$-tagging a challenging task.

\section{Detector scenarios}
\label{SEC:Scenarios}

A total of ten scenarios with different combinations of ideal detector components are studied. Each scenario is characterized by a different selection of the input features for the neural network which correspond to those properties of the jets that the detector components are designed to measure. For example, in a scenario with a tracking detector, the trajectory, the charge and the momentum of charged particles are used as input features. However, detector effects, like alignment, hit inefficiencies, track reconstruction, momentum resolution, charge misidentification, and hadron-collision-related effects such as pile-up or beam remnant are not taken into account. All of them are expected to reduce the performance of an $s$-tagger. In the following, all ten scenarios are briefly described and motivated. An overview of all scenarios as well as a summary of the input features is given in Table~\ref{TAB:scenarios}. All scenarios have in common that the jet $p_{\text{T}}$, $\eta$, and $\phi$ are always used as input to the neural network.

\begin{table}[h!]
\centering
\ras{0.9}
\begin{tabular}{lll}
\toprule
Name & Selection criteria &Input variables  \\ 
\toprule
Universal detector & $\tau>0$ & $E$, $\eta$, $\phi$, $m$ (4-momentum) \\ 
& & $r_0$, $\eta_0$, $\phi_0$ (origin)  \\ 
& & $q$ (charge) \\ 
& & $\tau$ (lifetime in lab system) \\
\midrule
Optimistic collider detector & $\tau>0$ & $E$, $\eta$, $\phi$, (4-momentum minus mass), \\ 
 & $r_f(\text{charged particle})  > 10\,\text{mm}$,& $r_i$, $\eta_i$, $\phi_i$ (initial measurement), \\ 
&  $r_f(\text{neutral particle})  > 1\,\text{m}$& $q$ (charge)\\
\toprule
Tracking detector & $\tau> 0$  & $p$, $\eta$, $\phi$ (4-momentum minus mass) \\
& $r_f > 10\,\text{mm}$ &$r_i$, $\eta_i$, $\phi_i$ (initial measurement)  \\
&$q \neq 0$& $\tau$ (lifetime in lab system) \\
&$r_0 < 1\,\text{m}$& $q$ (charge)\\
\midrule
Cherenkov detector& $\tau> 0$ & $p$, $\eta$, $\phi$, $m$ (4-momentum)  \\
 & $r_f > 10\,\text{mm}$ & $r_i$, $\eta_i$, $\phi_i$ (initial measurement) \\
& $q \neq 0$ & $\tau$ (lifetime in lab system) \\
&$r_0 < 1\,\text{m}$& $q$ (charge)\\
\toprule
Calorimeter without& $\tau>0$ & $E$, $\eta$, $\phi$ (3-momentum)\\
ECAL/HCAL separation& $r_0 < 1\,\text{m}$ & \\
& $r_f > 1\,\text{m}$ & \\
& no $\nu$ & \\
\midrule
Calorimeter with & $\tau>0$ & $E$, $\eta$, $\phi$ (3-momentum), \\
ECAL/HCAL separation  & $r_0 < 1\,\text{m}$ & particle category ($\gamma$/$e$, $\mu$, other) \\
& $r_f > 1\,\text{m}$ & \\
& no $\nu$ & \\
\bottomrule
\end{tabular} 
\caption{
List of all considered detector scenarios as a combination of ideal detector components. The second column shows the selection requirements imposed on the particles used as input to the neural networks, where $\tau$ means the lifetime of the particles, $r_0$ is the radial distance between the primary vertex and the point where the particle is created, and $r_f$ is the radial distance between the primary vertex and the decay vertex. The third column describes the variables that are used as inputs to the neural network. If the variable carries a subscript $0$, it refers to the spacepoint of creation, and if it carries a subscript $i$, it refers to the spacepoint of initial measurement.
 \label{TAB:scenarios}}
\end{table}

\subsection{Ideal detectors}

\subsubsection{Universal collider detector}

This scenario corresponds to an ideal detector that would be able to identify and measure the properties of all jet constituents with infinite precision. While this is of course unrealistic, this scenario helps to determine an upper bound on the performance that an $s$-tagger can possibly achieve.

All constituent particles of the jets are used as input to the recurrent neural network, independent of the position of their origin or decay inside the detector. No requirement on their momentum is imposed.
However, particles with zero lifetime are rejected because they do not enter the jet clustering algorithm. 

The neural network is provided with a set of nine input variables which fully describe the particles' properties: The particles' four-momenta are described by their energy $E$, pseudorapidity $\eta$, azimuthal angle $\phi$, and mass $m$. In addition, the linear trajectory of the particles is fully determined by their point of creation (either the primary vertex or secondary decay vertices defined by the radial distance from the primary vertex $r_0$ and the angles $\eta_0$ and $\phi_0$), as well as the direction of the four-momentum vector and the lifetime $\tau$. Finally, for an unambiguous identification of the particle type, the charge $q$ of the particles is used as an input variable.

\subsubsection{Optimistic Collider Detector}

A typical collider detector cannot measure particles arbitrarily close to the beam line. 
In order to take this aspect into account, an optimistic detector scenario is considered. In this scenario, the input to the neural network is modified by removing all  electrically charged particles that decay within a radius of $r = 10\,\text{mm}$. The value of the radius is inspired by the size of the beam pipe at the LHC experiments, around which tracking detectors are typically positioned. 
Because neutral particles are only detected in the calorimeter, all neutral particles decaying within a radius of $r = 1\,\text{m}$ are removed, which corresponds to the order of magnitude of the distance between the collision vertex and a typical calorimeter. This scenario is optimistic in the sense that the exact momentum or energy of the particles is measured, even if their signature in a realistic detector lacks information for a complete reconstruction. 
The input values of the particles points of creation ($r_0$, $\eta_0$, $\phi_0$) are altered to represent the point of initial detection. As a typical collider detector cannot reconstruct the particles' masses, they are removed as input variables as well. Similarly, the lifetime of the particles is not considered, as it can only be reconstructed for particles that decay inside a tracking detector.

\subsubsection{Tracking detector}

In most collider detectors, the tracking detector is the detector with the best spatial resolution. It measures so-called hits when charged particles pass through its active material, which is often structured in layers. The hits are then connected  using track reconstruction algorithms to reconstruct the charged particles' trajectories. The momentum of the particle and sign of its electric charge are determined from the curvature of the track in a magnetic field.

Since tracking detectors measure only charged particles, in the following scenarios, these are the only particles which are passed to the neural network  as input. Since tracking detectors measure the particles' three-momentum and electric charge\footnote{It is assumed that all particles carry either unit electric charge or are electrically neutral, which is true for all relevant leptons and hadrons. Hence, the charge is inferred from the measurement of the sign of the charge.}, the mass $m$ is not included as an input feature, and the particle momentum $p$ is used instead of the energy. Since this is an ideal tracking detector, no track reconstruction inefficiencies or resolutions for the momenta and angles are considered. 

The tracking detector is assumed to cover a cylindrical volume with radius $10\,\text{mm}<r<1\,\text{m}$. The point of initial detection ($r_0$, $\eta_0$, $\phi_0$) is modified as described previously for the optimistic collider detector. The lifetime ($\tau$) is modified to to represent the time observed in the instrumented volume\footnote{If the particle leaves the tracking detector without decaying, a large default value representing an ''infinite`` lifetime is used.}. If any particle decays within the radius of $r = 10\,\text{mm}$ or has its origin beyond a radius of $r = 1\,\text{m}$, it is removed from the list of input particles.

\subsubsection{Cherenkov detector}

In the past, Cherenkov detectors were used to identify charged kaons in order to discriminate $s$-jets from other jets~\cite{Kalelkar:2000ig,Abreu:1999cj}. Using the emission of Cherenkov light and the momentum measurement from the tracking detectors, Cherenkov detectors distinguish between charged particles with different masses. Similarly, the energy loss in a tracking detector ($\text{d}E/\text{d}x$) in combination with the measurement of the momentum can be used to determine the mass of a charged particle. Both of these methods typically have limitations for large momenta. However, in this idealized detector scenario, a detector that is able to perfectly measure the mass and three-momentum of charged particles is assumed. Similar to the tracking detector, the detector covers the space within a radius of $10\,\text{mm}<r<1\,\text{m}$. The only difference is that the particles' masses are used as an additional input feature.

\subsubsection{Calorimeters with and without separation into ECAL and HCAL}

Calorimeters measure the energy of electrons and photons via electromagnetic showers and the energy of both charged and neutral hadrons via hadronic showers. Typically, two separate types of calorimeters are used, one that is optimized for electromagnetic showers (ECAL) and one that is optimized for hadronic showers (HCAL).

\begin{figure}[t]
\centering
\includegraphics[width=.6\textwidth]{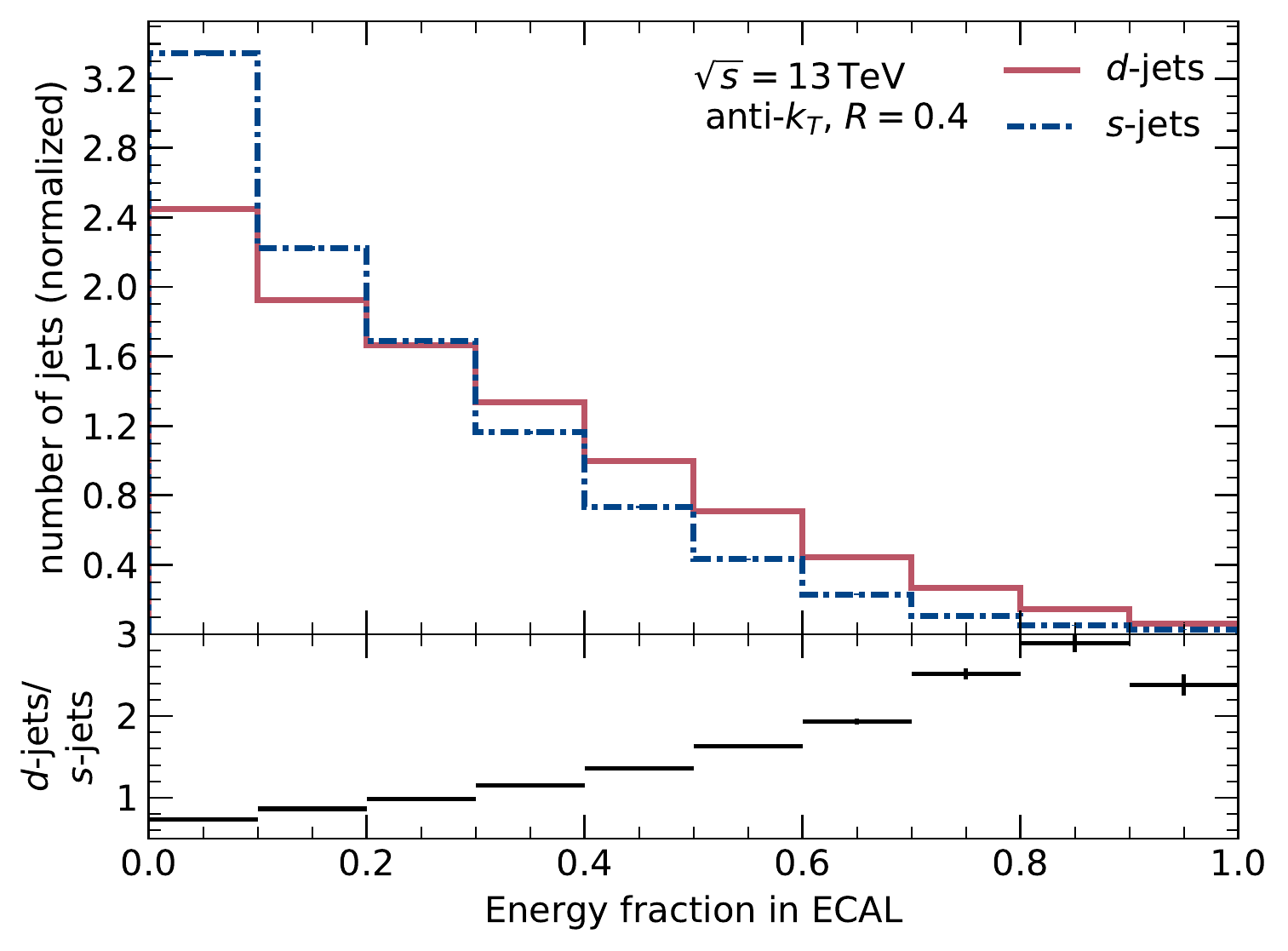} 
\caption{Distribution of the fraction of jet energy that is deposited in the ECAL, i.e. the fraction of energy carried by electrons and photons, for $s$-jets and $d$-jets.}
\label{FIG:EMFrac}
\end{figure}

One feature that can be used to discriminate between $s$- and $d$-jets is the fraction of energy that is deposited in the ECAL. Jets from down quarks tend to contain more photons from $\pi^0\rightarrow\gamma\gamma$ decays, while $s$-jets tend to contain more $K_L$ mesons, which typically do not decay before they reach the calorimeters. Figure~\ref{FIG:EMFrac} illustrates the fraction of energy of a jet carried by photons and electrons --- as would be measured by a perfect ECAL --- for both $s$-jets and $d$-jets at a radius of $r=1\,\text{m}$, where the distance of $1\,\text{m}$ from the primary vertex is inspired by calorimeters at the LHC. 

Moreover, the geometrical distribution of energy within the jets (substructure) could contain additional information for the discrimination of $s$-jets and $d$-jets. To evaluate the discrimination power of both, substructure and energy fraction carried by electrons and photons, two calorimeter scenarios are considered.

The first scenario is a calorimeter without separation into ECAL and HCAL, which can perfectly determine the energy and angles in the $\eta$--$\phi$ plane of a particle. The input to the neural network is the energy $E$, pseudorapidity $\eta$, and azimuthal angle $\phi$ at the radius $r=1\,\text{m}$ for all particles that pass through this radius, except for neutrinos $\nu$. This scenario is sensitive to the discrimination power of the jet substructure.
As those particles that reach the calorimeter initiate shower containing a large number of particles, it is assumed that the particle's decay chain is no longer resolvable inside the calorimeter.

A second calorimeter scenario passes an additional variable to the neural network which discriminates between three different particle categories: (1) electrons and photons, (2) muons, which are minimum ionizing particles that typically pass through the calorimeters and are then identified with muon detectors that are situated behind the calorimeters, and (3) all other particles, including foremost hadrons but also tau leptons. This emulates the classification of particles with a combination of ECAL and HCAL\footnote{In practice, the energy distributions in the ECAL and HCAL cannot be associated to individual particles.}. However, in contrast to real calorimeter setups, in this scenario, the ECAL and HCAL do not have any depth, which means that the neural network is not provided with the information of the longitudinal extension of the shower.
Any increase in separation power with respect to the first calorimeter scenario can be attributed to the separation of electromagnetic and hadronic energy deposits and --- to a lesser extent --- the presence of muons.

\subsection{Strange-hadron decays to two charged particles}
\label{SEC:neutral_to_charged}

It is possible to reconstruct decays of neutral strange-hadrons to two charged particles, such as $K_S \rightarrow \pi^+ \pi^-$ or $\Lambda^0 \rightarrow p \pi^-$, using a tracking detector. As such strange-hadrons are more prevalent in $s$-jets than in $d$-jets, it is worth asking how much discrimination power is contained in these $0\rightarrow+-$ decays.

To enhance the separation power of these  $0\rightarrow +-$ decays, the input to the tracking detector scenario is modified, replacing particles of opposite electric charge originating in the same vertex by a neutral particle whose kinematic properties are calculated from the sum of the 4-vectors of the assumed decay products.
In a cross check, only stable charged particles with trajectories within $\SI{10}{mm} < r < \SI{1}{m}$ are considered, therefore removing the discrimination power of $0\rightarrow+-$ decays.

\section{Neural-network architecture}

\begin{figure}

\centering
\noindent\tikzset{every picture/.style={line width=0.75pt}} 
\resizebox{0.7\textwidth}{0.62\textwidth}{      
\begin{tikzpicture}[x=0.75pt,y=0.75pt,yscale=-1,xscale=1]
\path (0,530); 
\clip (0,0) rectangle (600, 530);
\draw    (300,230) -- (380.42,325.08) ;
\draw    (270,230) -- (350.42,325.08) ;
\draw    (240,230) -- (320.42,325.08) ;
\draw    (210,230) -- (290.42,325.08) ;
\draw    (180,230) -- (260.42,325.08) ;
\draw    (150,230) -- (202.36,291.91) -- (230.42,325.08) ;
\draw    (430,230) -- (430,300) ;
\draw    (480,230) -- (480,300) ;
\draw    (530,230) -- (530,300) ;
\draw    (410,330.17) -- (390,420.17) ;
\draw    (440,330.17) -- (400,420.17) ;
\draw    (500,330.17) -- (420,420.17) ;
\draw    (470,330.17) -- (410,420.17) ;
\draw    (350,330.17) -- (370,420.17) ;
\draw    (320,330.17) -- (360,420.17) ;
\draw    (290,330.17) -- (350,420.17) ;
\draw    (260,330.17) -- (340,420.17) ;
\draw    (380.42,325.08) -- (380,415) ;
\draw  [fill={rgb, 255:red, 255; green, 255; blue, 255 }  ,fill opacity=1 ] (411.92,230) .. controls (411.92,220.01) and (420.01,211.92) .. (430,211.92) .. controls (439.99,211.92) and (448.08,220.01) .. (448.08,230) .. controls (448.08,239.99) and (439.99,248.08) .. (430,248.08) .. controls (420.01,248.08) and (411.92,239.99) .. (411.92,230) -- cycle ;
\draw  [fill={rgb, 255:red, 255; green, 255; blue, 255 }  ,fill opacity=1 ] (461.92,230) .. controls (461.92,220.01) and (470.01,211.92) .. (480,211.92) .. controls (489.99,211.92) and (498.08,220.01) .. (498.08,230) .. controls (498.08,239.99) and (489.99,248.08) .. (480,248.08) .. controls (470.01,248.08) and (461.92,239.99) .. (461.92,230) -- cycle ;
\draw  [fill={rgb, 255:red, 255; green, 255; blue, 255 }  ,fill opacity=1 ] (511.92,230) .. controls (511.92,220.01) and (520.01,211.92) .. (530,211.92) .. controls (539.99,211.92) and (548.08,220.01) .. (548.08,230) .. controls (548.08,239.99) and (539.99,248.08) .. (530,248.08) .. controls (520.01,248.08) and (511.92,239.99) .. (511.92,230) -- cycle ;
\draw  [fill={rgb, 255:red, 255; green, 255; blue, 255 }  ,fill opacity=1 ] (280,400.1) .. controls (280,394.61) and (284.45,390.17) .. (289.93,390.17) -- (470.07,390.17) .. controls (475.55,390.17) and (480,394.61) .. (480,400.1) -- (480,429.9) .. controls (480,435.39) and (475.55,439.83) .. (470.07,439.83) -- (289.93,439.83) .. controls (284.45,439.83) and (280,435.39) .. (280,429.9) -- cycle ;
\draw    (381.42,440.08) -- (381.92,498.25) ;
\draw    (401.42,440.08) -- (381.92,498.25) ;
\draw    (361.42,440.08) -- (381.92,498.25) ;
\draw    (421.42,440.08) -- (381.92,498.25) ;
\draw    (381.92,498.25) -- (341.42,440.08) ;
\draw    (319.5,440.08) -- (381.92,498.25) ;
\draw    (441.42,440.08) -- (381.92,498.25) ;
\draw    (290,2410) -- (412,2457.5) ;
\draw  [fill={rgb, 255:red, 255; green, 255; blue, 255 }  ,fill opacity=1 ] (210,310.03) .. controls (210,304.49) and (214.49,300) .. (220.03,300) -- (540.8,300) .. controls (546.34,300) and (550.83,304.49) .. (550.83,310.03) -- (550.83,340.13) .. controls (550.83,345.67) and (546.34,350.17) .. (540.8,350.17) -- (220.03,350.17) .. controls (214.49,350.17) and (210,345.67) .. (210,340.13) -- cycle ;
\draw  [fill={rgb, 255:red, 255; green, 255; blue, 255 }  ,fill opacity=1 ] (363.83,498.25) .. controls (363.83,488.26) and (371.93,480.17) .. (381.92,480.17) .. controls (391.9,480.17) and (400,488.26) .. (400,498.25) .. controls (400,508.24) and (391.9,516.33) .. (381.92,516.33) .. controls (371.93,516.33) and (363.83,508.24) .. (363.83,498.25) -- cycle ;
\draw  [color={rgb, 255:red, 128; green, 128; blue, 128 }  ,draw opacity=1 ][fill={rgb, 255:red, 255; green, 255; blue, 255 }  ,fill opacity=1 ] (79.17,106.03) .. controls (79.17,100.49) and (83.66,96) .. (89.2,96) -- (409.97,96) .. controls (415.51,96) and (420,100.49) .. (420,106.03) -- (420,136.13) .. controls (420,141.67) and (415.51,146.17) .. (409.97,146.17) -- (89.2,146.17) .. controls (83.66,146.17) and (79.17,141.67) .. (79.17,136.13) -- cycle ;
\draw  [color={rgb, 255:red, 128; green, 128; blue, 128 }  ,draw opacity=1 ][fill={rgb, 255:red, 255; green, 255; blue, 255 }  ,fill opacity=1 ] (81.08,38.08) .. controls (81.08,28.1) and (89.18,20) .. (99.17,20) .. controls (109.15,20) and (117.25,28.1) .. (117.25,38.08) .. controls (117.25,48.07) and (109.15,56.17) .. (99.17,56.17) .. controls (89.18,56.17) and (81.08,48.07) .. (81.08,38.08) -- cycle ;
\draw [color={rgb, 255:red, 128; green, 128; blue, 128 }  ,draw opacity=1 ]   (249.17,56.17) -- (249.17,96.17) ;
\draw [color={rgb, 255:red, 128; green, 128; blue, 128 }  ,draw opacity=1 ]   (299.17,56.17) -- (299.17,96.17) ;
\draw [color={rgb, 255:red, 128; green, 128; blue, 128 }  ,draw opacity=1 ]   (349.17,56.17) -- (349.17,96.17) ;
\draw [color={rgb, 255:red, 128; green, 128; blue, 128 }  ,draw opacity=1 ]   (199.17,56.17) -- (199.17,96.17) ;
\draw [color={rgb, 255:red, 128; green, 128; blue, 128 }  ,draw opacity=1 ]   (149.17,56.17) -- (149.17,96.17) ;
\draw [color={rgb, 255:red, 128; green, 128; blue, 128 }  ,draw opacity=1 ]   (99.17,56.17) -- (99.17,96.17) ;
\draw  [color={rgb, 255:red, 128; green, 128; blue, 128 }  ,draw opacity=1 ][fill={rgb, 255:red, 255; green, 255; blue, 255 }  ,fill opacity=1 ] (131.08,38.08) .. controls (131.08,28.1) and (139.18,20) .. (149.17,20) .. controls (159.15,20) and (167.25,28.1) .. (167.25,38.08) .. controls (167.25,48.07) and (159.15,56.17) .. (149.17,56.17) .. controls (139.18,56.17) and (131.08,48.07) .. (131.08,38.08) -- cycle ;
\draw  [color={rgb, 255:red, 128; green, 128; blue, 128 }  ,draw opacity=1 ][fill={rgb, 255:red, 255; green, 255; blue, 255 }  ,fill opacity=1 ] (181.08,38.08) .. controls (181.08,28.1) and (189.18,20) .. (199.17,20) .. controls (209.15,20) and (217.25,28.1) .. (217.25,38.08) .. controls (217.25,48.07) and (209.15,56.17) .. (199.17,56.17) .. controls (189.18,56.17) and (181.08,48.07) .. (181.08,38.08) -- cycle ;
\draw  [color={rgb, 255:red, 128; green, 128; blue, 128 }  ,draw opacity=1 ][fill={rgb, 255:red, 255; green, 255; blue, 255 }  ,fill opacity=1 ] (231.08,38.08) .. controls (231.08,28.1) and (239.18,20) .. (249.17,20) .. controls (259.15,20) and (267.25,28.1) .. (267.25,38.08) .. controls (267.25,48.07) and (259.15,56.17) .. (249.17,56.17) .. controls (239.18,56.17) and (231.08,48.07) .. (231.08,38.08) -- cycle ;
\draw  [color={rgb, 255:red, 128; green, 128; blue, 128 }  ,draw opacity=1 ][fill={rgb, 255:red, 255; green, 255; blue, 255 }  ,fill opacity=1 ] (281.08,38.08) .. controls (281.08,28.1) and (289.18,20) .. (299.17,20) .. controls (309.15,20) and (317.25,28.1) .. (317.25,38.08) .. controls (317.25,48.07) and (309.15,56.17) .. (299.17,56.17) .. controls (289.18,56.17) and (281.08,48.07) .. (281.08,38.08) -- cycle ;
\draw  [color={rgb, 255:red, 128; green, 128; blue, 128 }  ,draw opacity=1 ][fill={rgb, 255:red, 255; green, 255; blue, 255 }  ,fill opacity=1 ] (331.08,38.08) .. controls (331.08,28.1) and (339.18,20) .. (349.17,20) .. controls (359.15,20) and (367.25,28.1) .. (367.25,38.08) .. controls (367.25,48.07) and (359.15,56.17) .. (349.17,56.17) .. controls (339.18,56.17) and (331.08,48.07) .. (331.08,38.08) -- cycle ;
\draw  [color={rgb, 255:red, 128; green, 128; blue, 128 }  ,draw opacity=1 ][fill={rgb, 255:red, 255; green, 255; blue, 255 }  ,fill opacity=1 ] (381.08,38.08) .. controls (381.08,28.1) and (389.18,20) .. (399.17,20) .. controls (409.15,20) and (417.25,28.1) .. (417.25,38.08) .. controls (417.25,48.07) and (409.15,56.17) .. (399.17,56.17) .. controls (389.18,56.17) and (381.08,48.07) .. (381.08,38.08) -- cycle ;
\draw [color={rgb, 255:red, 128; green, 128; blue, 128 }  ,draw opacity=1 ]   (249.17,146.17) -- (249.17,186.17) ;
\draw [color={rgb, 255:red, 128; green, 128; blue, 128 }  ,draw opacity=1 ]   (289.17,146.17) -- (269.17,186.17) ;
\draw [color={rgb, 255:red, 128; green, 128; blue, 128 }  ,draw opacity=1 ]   (329.17,146.17) -- (289.17,186.17) ;
\draw [color={rgb, 255:red, 128; green, 128; blue, 128 }  ,draw opacity=1 ]   (369.17,146.17) -- (309.17,186.17) ;
\draw [color={rgb, 255:red, 128; green, 128; blue, 128 }  ,draw opacity=1 ]   (409.97,146.17) -- (329.17,186.17) ;
\draw [color={rgb, 255:red, 128; green, 128; blue, 128 }  ,draw opacity=1 ]   (209.17,146.17) -- (229.17,186.17) ;
\draw [color={rgb, 255:red, 128; green, 128; blue, 128 }  ,draw opacity=1 ]   (169.17,146.17) -- (209.17,186.17) ;
\draw [color={rgb, 255:red, 128; green, 128; blue, 128 }  ,draw opacity=1 ]   (129.17,146.17) -- (189.17,186.17) ;
\draw [color={rgb, 255:red, 128; green, 128; blue, 128 }  ,draw opacity=1 ]   (89.2,146.17) -- (169.17,186.17) ;
\draw  [color={rgb, 255:red, 128; green, 128; blue, 128 }  ,draw opacity=1 ][fill={rgb, 255:red, 255; green, 255; blue, 255 }  ,fill opacity=1 ] (149.17,196.1) .. controls (149.17,190.61) and (153.61,186.17) .. (159.1,186.17) -- (339.23,186.17) .. controls (344.72,186.17) and (349.17,190.61) .. (349.17,196.1) -- (349.17,225.9) .. controls (349.17,231.39) and (344.72,235.83) .. (339.23,235.83) -- (159.1,235.83) .. controls (153.61,235.83) and (149.17,231.39) .. (149.17,225.9) -- cycle ;
\draw [color={rgb, 255:red, 128; green, 128; blue, 128 }  ,draw opacity=1 ]   (399.17,56.17) -- (399.17,96.17) ;
\draw  [color={rgb, 255:red, 74; green, 74; blue, 74 }  ,draw opacity=1 ][fill={rgb, 255:red, 255; green, 255; blue, 255 }  ,fill opacity=1 ] (70,116.03) .. controls (70,110.49) and (74.49,106) .. (80.03,106) -- (400.8,106) .. controls (406.34,106) and (410.83,110.49) .. (410.83,116.03) -- (410.83,146.13) .. controls (410.83,151.67) and (406.34,156.17) .. (400.8,156.17) -- (80.03,156.17) .. controls (74.49,156.17) and (70,151.67) .. (70,146.13) -- cycle ;
\draw  [color={rgb, 255:red, 74; green, 74; blue, 74 }  ,draw opacity=1 ][fill={rgb, 255:red, 255; green, 255; blue, 255 }  ,fill opacity=1 ] (71.92,48.08) .. controls (71.92,38.1) and (80.01,30) .. (90,30) .. controls (99.99,30) and (108.08,38.1) .. (108.08,48.08) .. controls (108.08,58.07) and (99.99,66.17) .. (90,66.17) .. controls (80.01,66.17) and (71.92,58.07) .. (71.92,48.08) -- cycle ;
\draw [color={rgb, 255:red, 74; green, 74; blue, 74 }  ,draw opacity=1 ]   (240,66.17) -- (240,106.17) ;
\draw [color={rgb, 255:red, 74; green, 74; blue, 74 }  ,draw opacity=1 ]   (290,66.17) -- (290,106.17) ;
\draw [color={rgb, 255:red, 74; green, 74; blue, 74 }  ,draw opacity=1 ]   (340,66.17) -- (340,106.17) ;
\draw [color={rgb, 255:red, 74; green, 74; blue, 74 }  ,draw opacity=1 ]   (190,66.17) -- (190,106.17) ;
\draw [color={rgb, 255:red, 74; green, 74; blue, 74 }  ,draw opacity=1 ]   (140,66.17) -- (140,106.17) ;
\draw [color={rgb, 255:red, 74; green, 74; blue, 74 }  ,draw opacity=1 ]   (90,66.17) -- (90,106.17) ;
\draw  [color={rgb, 255:red, 74; green, 74; blue, 74 }  ,draw opacity=1 ][fill={rgb, 255:red, 255; green, 255; blue, 255 }  ,fill opacity=1 ] (121.92,48.08) .. controls (121.92,38.1) and (130.01,30) .. (140,30) .. controls (149.99,30) and (158.08,38.1) .. (158.08,48.08) .. controls (158.08,58.07) and (149.99,66.17) .. (140,66.17) .. controls (130.01,66.17) and (121.92,58.07) .. (121.92,48.08) -- cycle ;
\draw  [color={rgb, 255:red, 74; green, 74; blue, 74 }  ,draw opacity=1 ][fill={rgb, 255:red, 255; green, 255; blue, 255 }  ,fill opacity=1 ] (171.92,48.08) .. controls (171.92,38.1) and (180.01,30) .. (190,30) .. controls (199.99,30) and (208.08,38.1) .. (208.08,48.08) .. controls (208.08,58.07) and (199.99,66.17) .. (190,66.17) .. controls (180.01,66.17) and (171.92,58.07) .. (171.92,48.08) -- cycle ;
\draw  [color={rgb, 255:red, 74; green, 74; blue, 74 }  ,draw opacity=1 ][fill={rgb, 255:red, 255; green, 255; blue, 255 }  ,fill opacity=1 ] (221.92,48.08) .. controls (221.92,38.1) and (230.01,30) .. (240,30) .. controls (249.99,30) and (258.08,38.1) .. (258.08,48.08) .. controls (258.08,58.07) and (249.99,66.17) .. (240,66.17) .. controls (230.01,66.17) and (221.92,58.07) .. (221.92,48.08) -- cycle ;
\draw  [color={rgb, 255:red, 74; green, 74; blue, 74 }  ,draw opacity=1 ][fill={rgb, 255:red, 255; green, 255; blue, 255 }  ,fill opacity=1 ] (271.92,48.08) .. controls (271.92,38.1) and (280.01,30) .. (290,30) .. controls (299.99,30) and (308.08,38.1) .. (308.08,48.08) .. controls (308.08,58.07) and (299.99,66.17) .. (290,66.17) .. controls (280.01,66.17) and (271.92,58.07) .. (271.92,48.08) -- cycle ;
\draw  [color={rgb, 255:red, 74; green, 74; blue, 74 }  ,draw opacity=1 ][fill={rgb, 255:red, 255; green, 255; blue, 255 }  ,fill opacity=1 ] (321.92,48.08) .. controls (321.92,38.1) and (330.01,30) .. (340,30) .. controls (349.99,30) and (358.08,38.1) .. (358.08,48.08) .. controls (358.08,58.07) and (349.99,66.17) .. (340,66.17) .. controls (330.01,66.17) and (321.92,58.07) .. (321.92,48.08) -- cycle ;
\draw  [color={rgb, 255:red, 74; green, 74; blue, 74 }  ,draw opacity=1 ][fill={rgb, 255:red, 255; green, 255; blue, 255 }  ,fill opacity=1 ] (371.92,48.08) .. controls (371.92,38.1) and (380.01,30) .. (390,30) .. controls (399.99,30) and (408.08,38.1) .. (408.08,48.08) .. controls (408.08,58.07) and (399.99,66.17) .. (390,66.17) .. controls (380.01,66.17) and (371.92,58.07) .. (371.92,48.08) -- cycle ;
\draw [color={rgb, 255:red, 74; green, 74; blue, 74 }  ,draw opacity=1 ]   (240,156.17) -- (240,196.17) ;
\draw [color={rgb, 255:red, 74; green, 74; blue, 74 }  ,draw opacity=1 ]   (280,156.17) -- (260,196.17) ;
\draw [color={rgb, 255:red, 74; green, 74; blue, 74 }  ,draw opacity=1 ]   (320,156.17) -- (280,196.17) ;
\draw [color={rgb, 255:red, 74; green, 74; blue, 74 }  ,draw opacity=1 ]   (360,156.17) -- (300,196.17) ;
\draw [color={rgb, 255:red, 74; green, 74; blue, 74 }  ,draw opacity=1 ]   (400.8,156.17) -- (320,196.17) ;
\draw [color={rgb, 255:red, 74; green, 74; blue, 74 }  ,draw opacity=1 ]   (200,156.17) -- (220,196.17) ;
\draw [color={rgb, 255:red, 74; green, 74; blue, 74 }  ,draw opacity=1 ]   (160,156.17) -- (200,196.17) ;
\draw [color={rgb, 255:red, 74; green, 74; blue, 74 }  ,draw opacity=1 ]   (120,156.17) -- (180,196.17) ;
\draw [color={rgb, 255:red, 74; green, 74; blue, 74 }  ,draw opacity=1 ]   (80.03,156.17) -- (160,196.17) ;
\draw  [color={rgb, 255:red, 74; green, 74; blue, 74 }  ,draw opacity=1 ][fill={rgb, 255:red, 255; green, 255; blue, 255 }  ,fill opacity=1 ] (140,206.1) .. controls (140,200.61) and (144.45,196.17) .. (149.93,196.17) -- (330.07,196.17) .. controls (335.55,196.17) and (340,200.61) .. (340,206.1) -- (340,235.9) .. controls (340,241.39) and (335.55,245.83) .. (330.07,245.83) -- (149.93,245.83) .. controls (144.45,245.83) and (140,241.39) .. (140,235.9) -- cycle ;
\draw [color={rgb, 255:red, 74; green, 74; blue, 74 }  ,draw opacity=1 ]   (390,66.17) -- (390,106.17) ;
\draw  [fill={rgb, 255:red, 255; green, 255; blue, 255 }  ,fill opacity=1 ] (59.17,126.03) .. controls (59.17,120.49) and (63.66,116) .. (69.2,116) -- (389.97,116) .. controls (395.51,116) and (400,120.49) .. (400,126.03) -- (400,156.13) .. controls (400,161.67) and (395.51,166.17) .. (389.97,166.17) -- (69.2,166.17) .. controls (63.66,166.17) and (59.17,161.67) .. (59.17,156.13) -- cycle ;
\draw  [fill={rgb, 255:red, 255; green, 255; blue, 255 }  ,fill opacity=1 ] (61.08,58.08) .. controls (61.08,48.1) and (69.18,40) .. (79.17,40) .. controls (89.15,40) and (97.25,48.1) .. (97.25,58.08) .. controls (97.25,68.07) and (89.15,76.17) .. (79.17,76.17) .. controls (69.18,76.17) and (61.08,68.07) .. (61.08,58.08) -- cycle ;
\draw    (229.17,76.17) -- (229.17,116.17) ;
\draw    (279.17,76.17) -- (279.17,116.17) ;
\draw    (329.17,76.17) -- (329.17,116.17) ;
\draw    (179.17,76.17) -- (179.17,116.17) ;
\draw    (129.17,76.17) -- (129.17,116.17) ;
\draw    (79.17,76.17) -- (79.17,116.17) ;
\draw  [fill={rgb, 255:red, 255; green, 255; blue, 255 }  ,fill opacity=1 ] (111.08,58.08) .. controls (111.08,48.1) and (119.18,40) .. (129.17,40) .. controls (139.15,40) and (147.25,48.1) .. (147.25,58.08) .. controls (147.25,68.07) and (139.15,76.17) .. (129.17,76.17) .. controls (119.18,76.17) and (111.08,68.07) .. (111.08,58.08) -- cycle ;
\draw  [fill={rgb, 255:red, 255; green, 255; blue, 255 }  ,fill opacity=1 ] (161.08,58.08) .. controls (161.08,48.1) and (169.18,40) .. (179.17,40) .. controls (189.15,40) and (197.25,48.1) .. (197.25,58.08) .. controls (197.25,68.07) and (189.15,76.17) .. (179.17,76.17) .. controls (169.18,76.17) and (161.08,68.07) .. (161.08,58.08) -- cycle ;
\draw  [fill={rgb, 255:red, 255; green, 255; blue, 255 }  ,fill opacity=1 ] (211.08,58.08) .. controls (211.08,48.1) and (219.18,40) .. (229.17,40) .. controls (239.15,40) and (247.25,48.1) .. (247.25,58.08) .. controls (247.25,68.07) and (239.15,76.17) .. (229.17,76.17) .. controls (219.18,76.17) and (211.08,68.07) .. (211.08,58.08) -- cycle ;
\draw  [fill={rgb, 255:red, 255; green, 255; blue, 255 }  ,fill opacity=1 ] (261.08,58.08) .. controls (261.08,48.1) and (269.18,40) .. (279.17,40) .. controls (289.15,40) and (297.25,48.1) .. (297.25,58.08) .. controls (297.25,68.07) and (289.15,76.17) .. (279.17,76.17) .. controls (269.18,76.17) and (261.08,68.07) .. (261.08,58.08) -- cycle ;
\draw  [fill={rgb, 255:red, 255; green, 255; blue, 255 }  ,fill opacity=1 ] (311.08,58.08) .. controls (311.08,48.1) and (319.18,40) .. (329.17,40) .. controls (339.15,40) and (347.25,48.1) .. (347.25,58.08) .. controls (347.25,68.07) and (339.15,76.17) .. (329.17,76.17) .. controls (319.18,76.17) and (311.08,68.07) .. (311.08,58.08) -- cycle ;
\draw  [fill={rgb, 255:red, 255; green, 255; blue, 255 }  ,fill opacity=1 ] (361.08,58.08) .. controls (361.08,48.1) and (369.18,40) .. (379.17,40) .. controls (389.15,40) and (397.25,48.1) .. (397.25,58.08) .. controls (397.25,68.07) and (389.15,76.17) .. (379.17,76.17) .. controls (369.18,76.17) and (361.08,68.07) .. (361.08,58.08) -- cycle ;
\draw    (229.17,166.17) -- (229.17,206.17) ;
\draw    (269.17,166.17) -- (249.17,206.17) ;
\draw    (309.17,166.17) -- (269.17,206.17) ;
\draw    (349.17,166.17) -- (289.17,206.17) ;
\draw    (389.97,166.17) -- (309.17,206.17) ;
\draw    (189.17,166.17) -- (209.17,206.17) ;
\draw    (149.17,166.17) -- (189.17,206.17) ;
\draw    (109.17,166.17) -- (169.17,206.17) ;
\draw    (69.2,166.17) -- (149.17,206.17) ;
\draw  [fill={rgb, 255:red, 255; green, 255; blue, 255 }  ,fill opacity=1 ] (129.17,216.1) .. controls (129.17,210.61) and (133.61,206.17) .. (139.1,206.17) -- (319.23,206.17) .. controls (324.72,206.17) and (329.17,210.61) .. (329.17,216.1) -- (329.17,245.9) .. controls (329.17,251.39) and (324.72,255.83) .. (319.23,255.83) -- (139.1,255.83) .. controls (133.61,255.83) and (129.17,251.39) .. (129.17,245.9) -- cycle ;
\draw    (379.17,76.17) -- (379.17,116.17) ;

\draw (310,320) node [anchor=north west][inner sep=0.75pt]   [align=left] {{\large dense layer, 64 nodes}};
\draw (310,408) node [anchor=north west][inner sep=0.75pt]   [align=left] {{\large dense layer, 32 nodes}};
\draw (433,22) node [anchor=north west][inner sep=0.75pt]   [align=left] {{\large input nodes for }\\{\large constituent variables}};
\draw (412,187) node [anchor=north west][inner sep=0.75pt]   [align=left] {{\large input nodes for jet variables}};
\draw (155,133) node [anchor=north west][inner sep=0.75pt]   [align=left] {{\large LSTM layer, 64 nodes}};
\draw (155,223) node [anchor=north west][inner sep=0.75pt]   [align=left] {{\large LSTM layer, 32 nodes}};
\end{tikzpicture}}
\caption{Visualisation of the neural network architecture used. Circles visualise input and output nodes, while boxes illustrate both LSTM and dense layers. The number of connections between layers illustrated as lines does not represent the number of connections.\label{FIG:architecture}}
\end{figure}
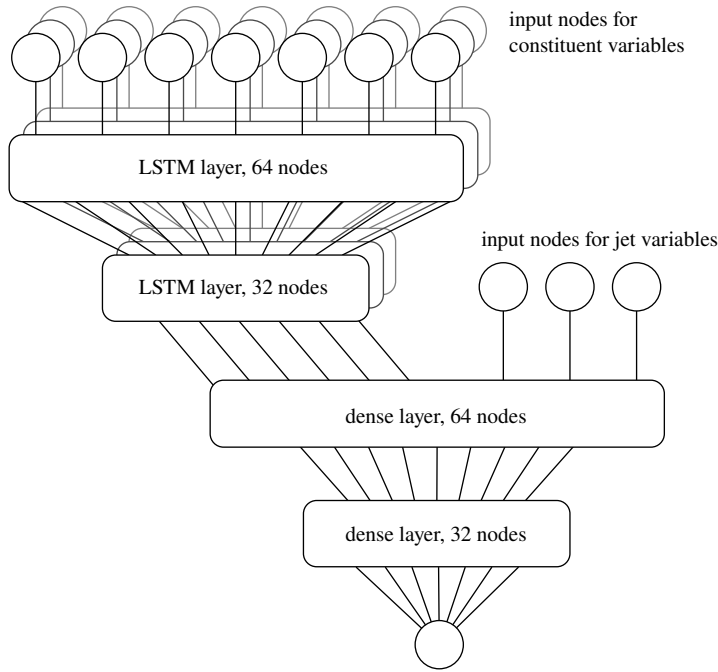

The neural network consists of a combination of LSTM layers and fully-connected feed-forward layers. The use of LSTM layers is motivated by two aspects. First, recurrent neural-networks are a possible way of handling the variable number of constituent particles in a jet. Second, LSTMs are capable of correlating both local clusters of information (such as information about a particle decay) as well as global features (such as overall jet momentum), which is a crucial property when the aim is to resolve the overall shower development.

For each scenario, the input features as discussed in the previous section are used. All input features are preprocessed ensuring that their values are in a comparable range using scikit-learn's~\cite{Pedregosa:2012toh} RobustScaler, which centers the distributions around zero and divides them by the range between their first and third quartiles.
The full data set of jets is split into a training sample (3/5), a validation sample (1/5) to monitor the neural network performance during the training and hyperparameter optimization, and an independent test sample (1/5) to evaluate the performance after the training.

The neural networks are implemented with Keras \cite{chollet2015keras} using TensorFlow \cite{tensorflow2015-whitepaper} as backend. 
To evaluate their performance, modified receiver operating characteristic (ROC) curves are used, which show the relation between signal efficiency ($\varepsilon_{s\text{-jets}}$) and background rejection ($1 / \varepsilon_{d\text{-jets}}$) for any cut on the output distribution of the neural network. As performance metric, the area under the curve (AUC) of the ROC curve is used, which can take on any value between 0.5 (no separation) and 1.0 (perfect separation) and is calculated with scikit-learn's~\cite{Pedregosa:2012toh} AUC score.

Figure~\ref{FIG:architecture} illustrates the neural networks' architecture. Their input layers contain as many nodes as there are constituent-level input features used in the respective scenario times the maximum number of considered particles (50). For example, the neural network for the universal-collider-detector scenario has $3\times50$ input nodes, while the neural network for the calorimeter scenario without separation into electromagnetic and hadronic components has $9\times50$ input nodes. In case there are less than 50 particles, a mask is applied in order to remove unused nodes, which makes the input dimension effectively smaller. 
The remaining part of all neural networks are structured in the same way for all scenarios and consist of two LSTM layers with 64 and 32 units followed by two fully-connected feed-forward (dense) layers with 64 and 32 nodes and a single output node. In addition to the 32 output nodes from the second LSTM layer, the $p_{\mathrm{T}}$, $\eta$, and $\phi$ of the jet are fed to the first dense layer.

\begin{figure}
\centering
\includegraphics[width=0.6\textwidth]{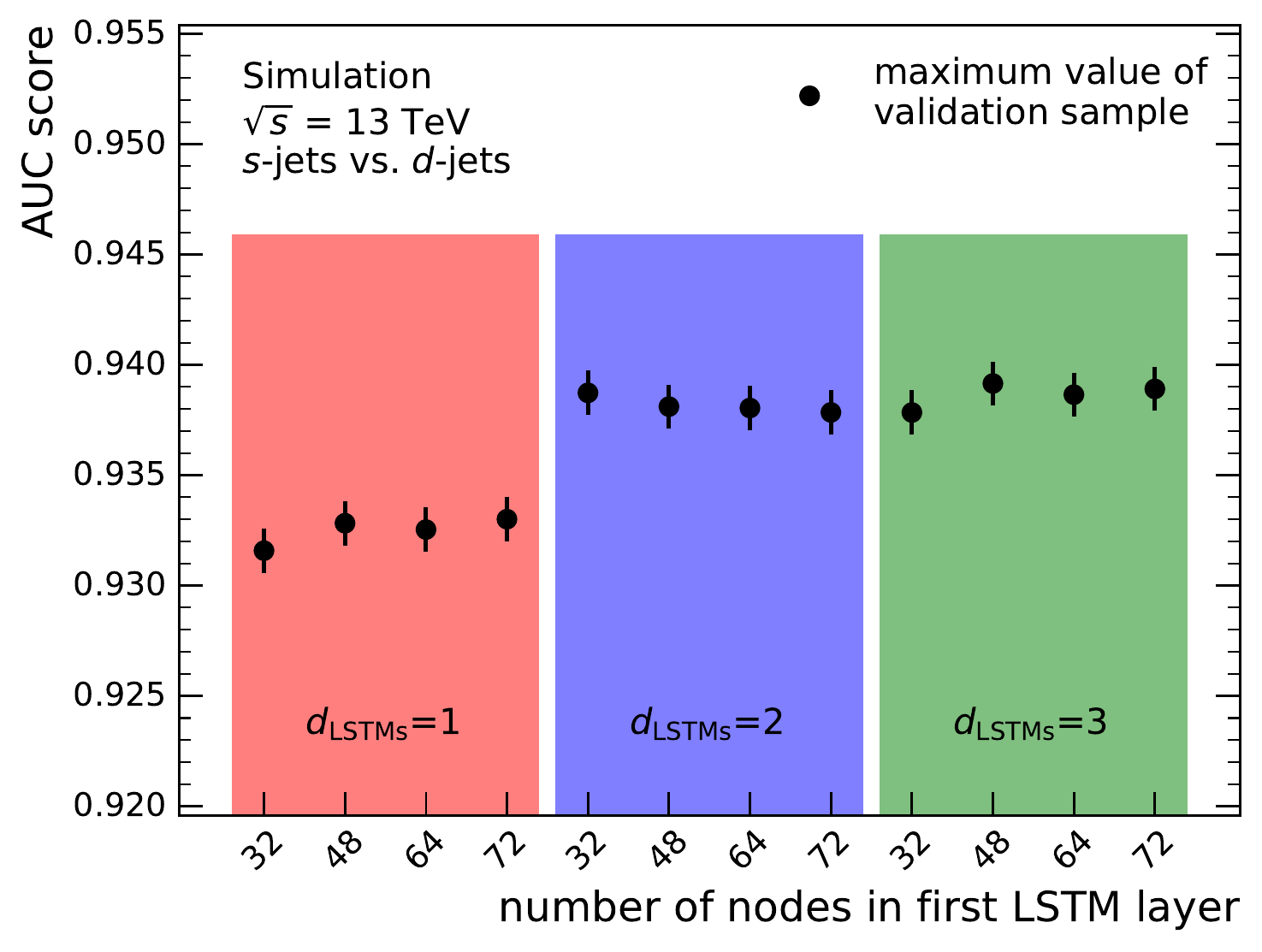}
\caption{Comparison of the AUC of the neural network trained using all particles as input for different neural network structures. Both the number of LSTM layers ($d_\text{LSTMs}$) and the number of nodes in the LSTM layers are varied. The number of nodes shown is the number of nodes in the first LSTM layer and all following LSTM layers contain half as many nodes as the previous layer. The structure of the dense layers is not varied for the comparison shown. The AUC scores are evaluated on the validation sample and the uncertainties correspond to the statistical uncertainty of the sample.\label{FIG:AUC_comp}}
\end{figure}

This configuration ensures that the capacity of the neural network is large enough in all scenarios. As the structure of the LSTM layers has the largest impact on the effective capacity, Figure~\ref{FIG:AUC_comp} shows the AUC score of different LSTM layer configurations. For simplicity, each LSTM layer is set to contain half as many nodes as the previous LSTM layer. While the AUC score increases when adding a second LSTM layer, any further addition of LSTM layers does not lead to a significant improvement in performance. The performance is stable under the variation of the number of nodes in the first and all subsequent LSTM layers while keeping the ratio of nodes in two consecutive layers constant. 

An $L_2$-norm penalty term is applied to weights of recurrent connections and to weights connecting each hidden layer in order to regularize the network. The same regularization strength is applied to each layer, but the value is optimized for each of the detector scenarios in order to find the optimal effective capacity by minimizing overtraining and maximizing the performance.

In both LSTM layers, the hyperbolic tangent function is used as activation for the input and output gate, while all gates are regulated using the sigmoid function. 
All nodes of the dense hidden layers are activated using the rectified linear unit, while the output node is activated using the sigmoid function.

During the training process, the binary cross entropy is optimized by the Adam algorithm~\cite{Kingma:2014vow} with an initial learning rate of 0.001 and a batch size of 1024 jets.

As an additional regularization measure, early stopping is used, which is triggered if the AUC value of the validation set has not improved by more than 0.0001 over the course of 20 epochs. The neural network model of the epoch with the largest AUC value  on the validation set is used to evaluate the final performance of the neural network.

\section{Results}

In the following, the classification performances of all ideal detector scenarios are discussed and compared. The comparison is divided into four different subsections: A comparison of (1) universal and optimistic collider detectors, (2) tracking and Cherenkov detectors, (3) both calorimeter detectors, and (4) a scenario with and without explicit reconstruction of $K_S\rightarrow\pi^+\pi^-$ decays.

\subsection{Collider detectors}

\begin{figure}[t]
\centering
\includegraphics[width=.6\textwidth]{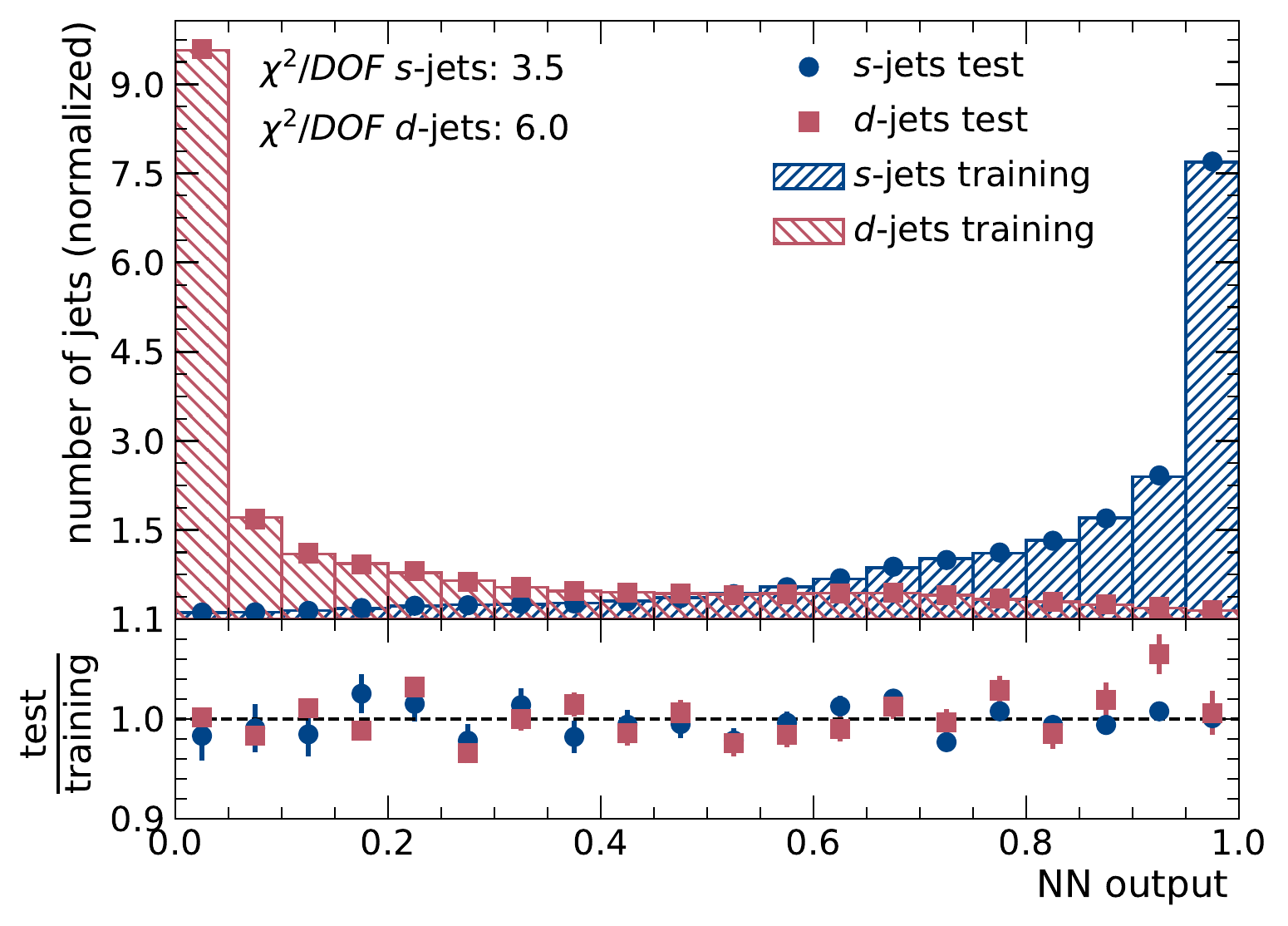}
\caption{Output distribution of the neural network in the universal collider detector scenario. The output for the training sample as well as for the independent test sample is compared for $s$-jets and $d$-jets.}
\label{FIG:all}
\end{figure}

Figure~\ref{FIG:all} shows the output distributions of the neural network trained in the scenario of the universal collider detector. $s$-jets and $d$-jets are separated well, forming two distinct peaks around values of 0 for $d$-jets and 1 for $s$-jets, which means many jets can be clearly identified correctly as $d$-jets and $s$-jets. While the $\chi^2$ per degree of freedom between the output distribution of the training and the test sample is relatively large (3.5 for $s$-jets and 6.0 for $d$-jets), their comparison shows that there is no indication of strong overtraining as there is no clear bias visible in the ratio between the test and training data set. 

\begin{figure}[t]
\centering
\includegraphics[width=.6\textwidth]{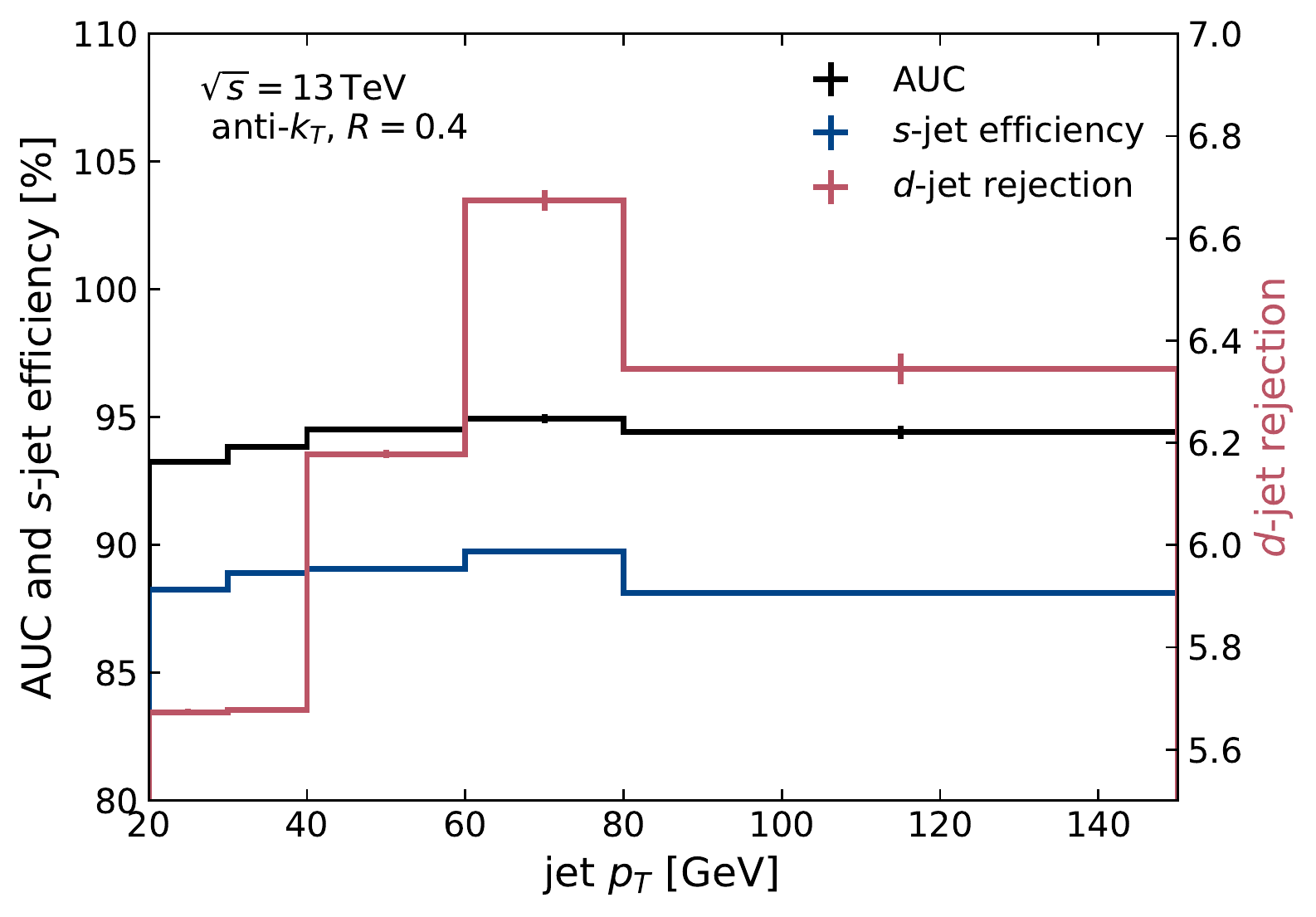}
\caption{The AUC, $s$-jet efficiency, and $d$-jet rejection for different jet $p_{\mathrm{T}}$ considering the neural network scenario trained using all particles. The efficiency and rejection are calculated in the test sample requiring a neural network output value larger 0.5 for a jet to pass the selection. All uncertainties are statistical uncertainties of the test sample.}
\label{FIG:all_pt}
\end{figure}

Figure~\ref{FIG:all_pt} illustrates the $p_{\mathrm{T}}$-dependence of the classification. The AUC as well as the $s$-jet efficiency and the $d$-jet rejection given a cut on the neural network output value of 0.5 increase up to a jet $p_{\mathrm{T}}=\SI{60}{GeV}$ as the number of constituent particles increases. Their decrease for $p_{\mathrm{T}}>\SI{60}{GeV}$ can be attributed to a larger amount of pions, which increases the similarity of $s$- and $d$-jets with respect to their net strangeness.

\begin{figure}[t]
\centering
\includegraphics[width=.6\textwidth]{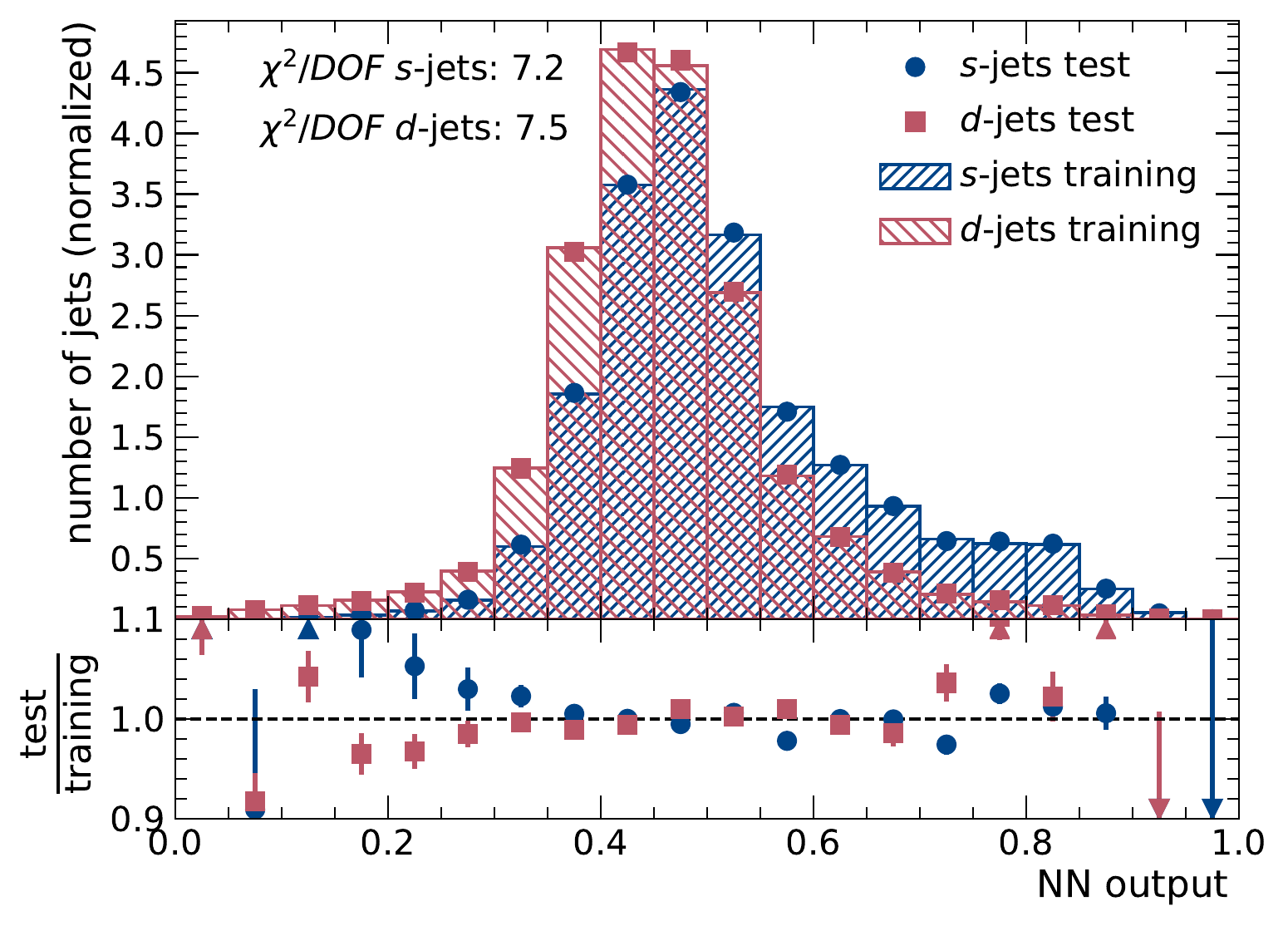}
\caption{Output distribution of the neural network in the optimistic collider detector scenario. The output for the training sample as well as for the independent test sample is compared for $s$-jets and $d$-jets.}
\label{FIG:reco}
\end{figure}

\begin{figure}[t]
\centering
\includegraphics[width=.6\textwidth]{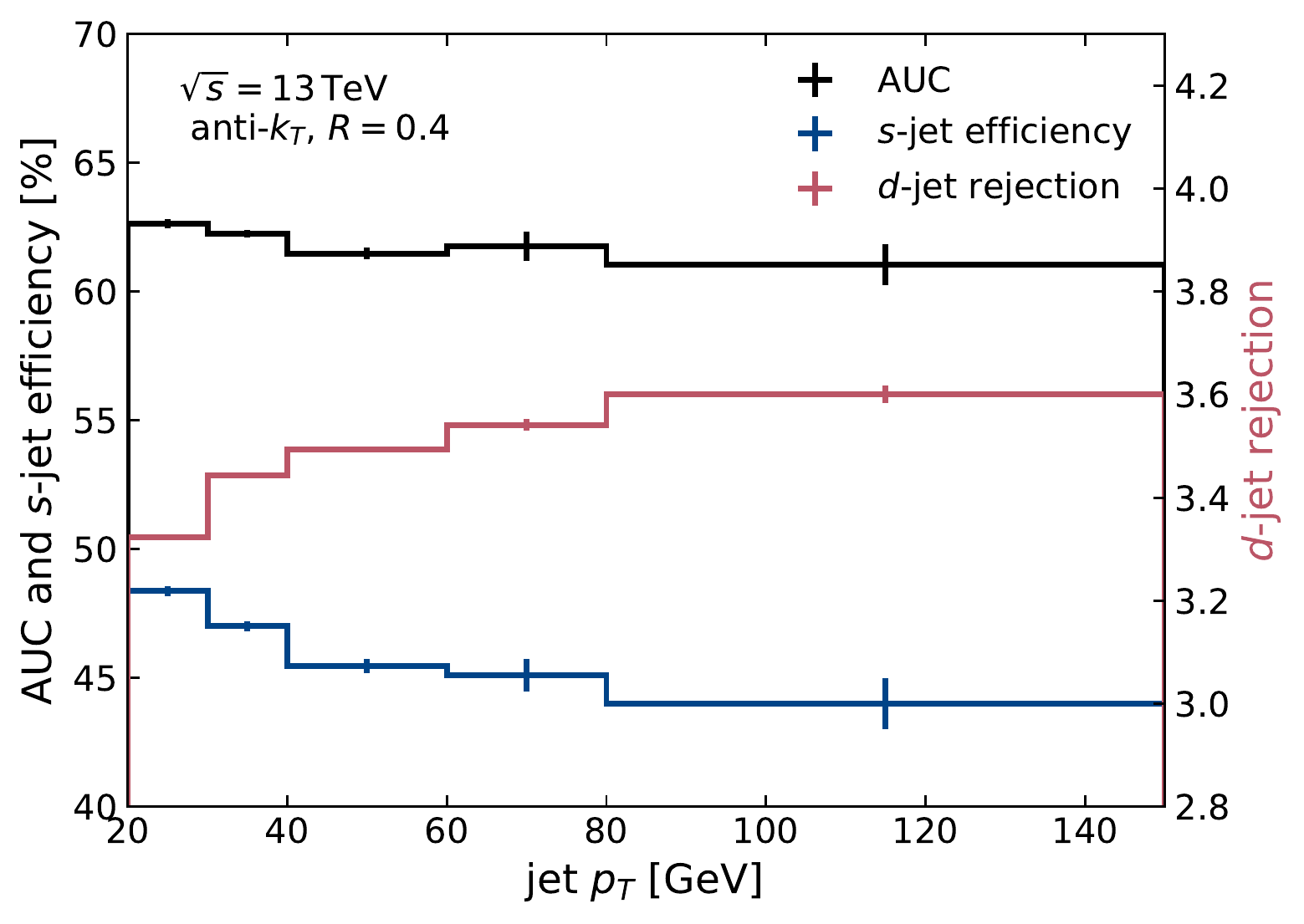}
\caption{AUC, $s$-jet efficiency, and $d$-jet rejection for different jet $p_{\mathrm{T}}$ considering the optimistic collider detector scenario. The efficiency and rejection are calculated in the test sample requiring a neural network output value larger 0.5 for a jet to pass the selection. All uncertainties are statistical uncertainties of the test sample.}
\label{FIG:reco_pt}
\end{figure}

As a real collider detector can neither detect particles before they reach the tracker or the calorimeter, respectively, nor determine their mass, a more stringent limit on the maximal performance of an $s$-tagger is derived in the optimistic collider detector scenario. Figure~\ref{FIG:reco} shows the output of the corresponding neural network. No significant overtraining is observed in the ratio between test and training data. The separation between $s$-jets and $d$-jets is significantly smaller as simple particle identification by using the mass is not possible  and neutral particles decaying within a radius of $r<\SI{1}{m}$ such as $\pi^0$, $K_S$, and $\Lambda^0$  are removed. 
Figure~\ref{FIG:reco_pt} shows the $s$-jet efficiency and $d$-jet rejection of this scenario for different $p_{\mathrm{T}}$ given a cut on the neural network output value of 0.5, as well as the AUC score for different $p_{\mathrm{T}}$.  The $s$-jet efficiency decreases and the $d$-jet rejection increases for larger $p_{\mathrm{T}}$. Overall, $s$- and $d$-jets become slightly more alike for larger momenta because of an increasing number of constituent particles and therefore an increased number of kaons.

\begin{figure}[t]
\centering
\includegraphics[width=.6\textwidth]{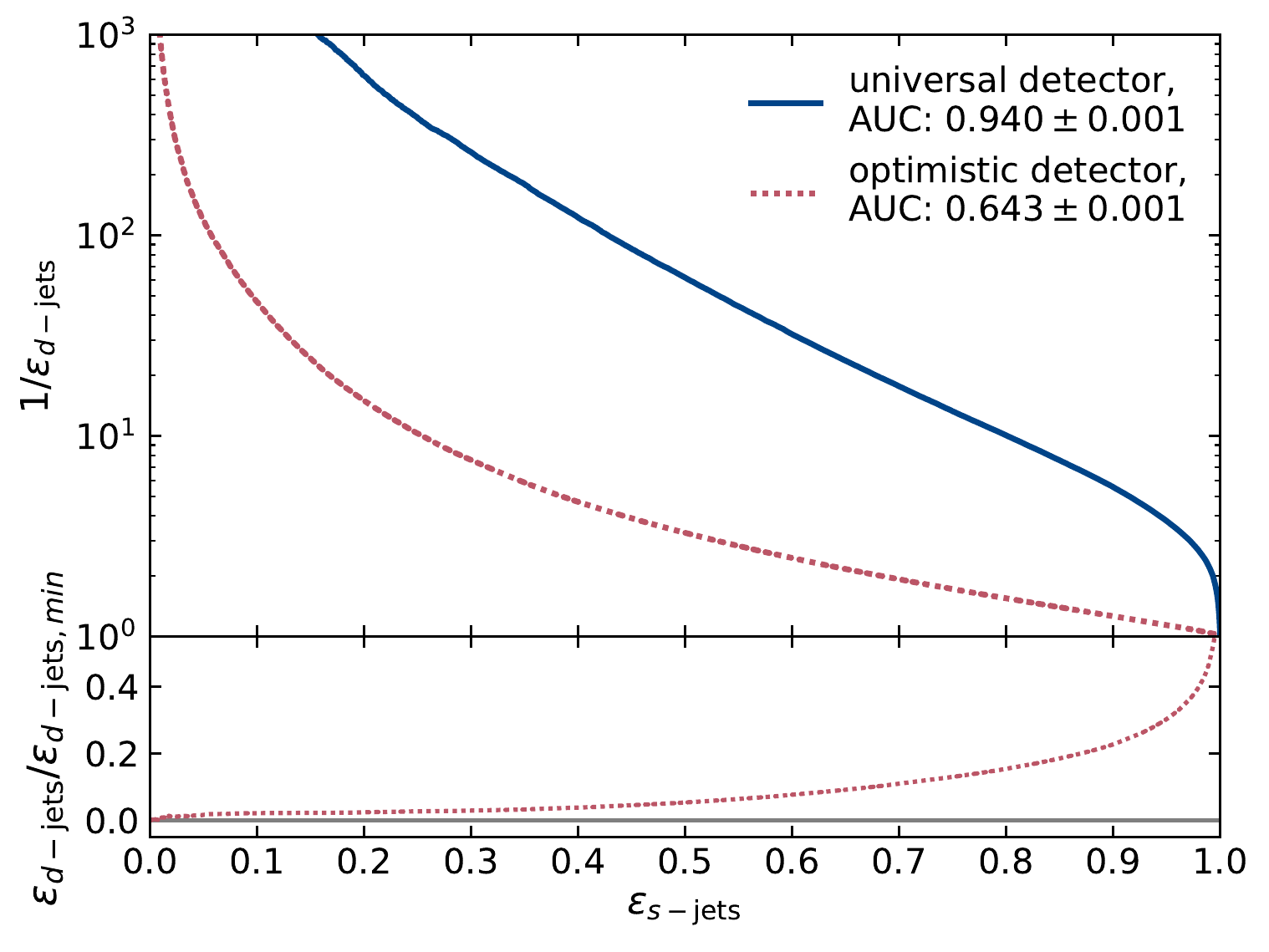}
\caption{ROC curves illustrating the classification power of neural networks of the universal detector scenario and the optimistic detector scenario that includes electrically charged particles detectable at radii $\SI{1}{mm} < r < \SI{1}{m}$ and neutral particles detectable at $r=\SI{1}{m}$. The signal is composed of $s$-jets, and the background is composed of $d$-jets. The ratio beneath the ROC curves shows the efficiency of $d$-jets in the universal detector scenario divided by the efficiency of $d$-jets in the optimistic detector scenario.
The efficiencies are evaluated on the test sample and the uncertainty in the area under the curve is the statistical uncertainty associated with that sample.}
\label{FIG:ROC_all}
\end{figure}

Figure~\ref{FIG:ROC_all} compares the ROC curves of the universal detector and the collider detector scenario as evaluated on the test sample. The statistical uncertainty of the AUC scores is derived via bootstrapping: Samples of the same size as the test sample are generated by drawing from it with replacement. As uncertainty, the standard deviation of the AUC score for 100 of such samples is used. 
With an AUC of $0.940\pm 0.001$, the universal detector scenario shows good separation between $s$- and $d$-jets, underlining that the particle content of $s$- and $d$-jet is quite different. It was found that the information in soft jet constituents with a $p_{\mathrm{T}} <\SI{5}{GeV}$ have no large impact on the separation power, as removing them reduces the AUC score of the universal collider detector scenario to 0.936. 
However, in the optimistic collider detector scenario, the AUC drops to $0.643\pm 0.001$.

This means that, while the hadronisation patterns of $s$- and $d$-jets differ, lack of particle identification and reconstructability of neutral particles significantly reduces the performance of an $s$-tagger. 

Ref.~\cite{Nakai:2020kuu} contains similar scenarios considering a combination of tracking detectors and calorimeters. In a scenario based on truth information, a separation with an AUC value of 0.67 \footnote{Differences between the separation power achieved in this paper and Ref.~\cite{Nakai:2020kuu} can be attributed to different jet kinematics.} is achieved, while a reduced separation with an AUC value of 0.64 is obtained for a more realistic detector scenario.

\subsection{Tracking and Cherenkov detectors}

\begin{figure}[t]
\centering
\includegraphics[width=.6\textwidth]{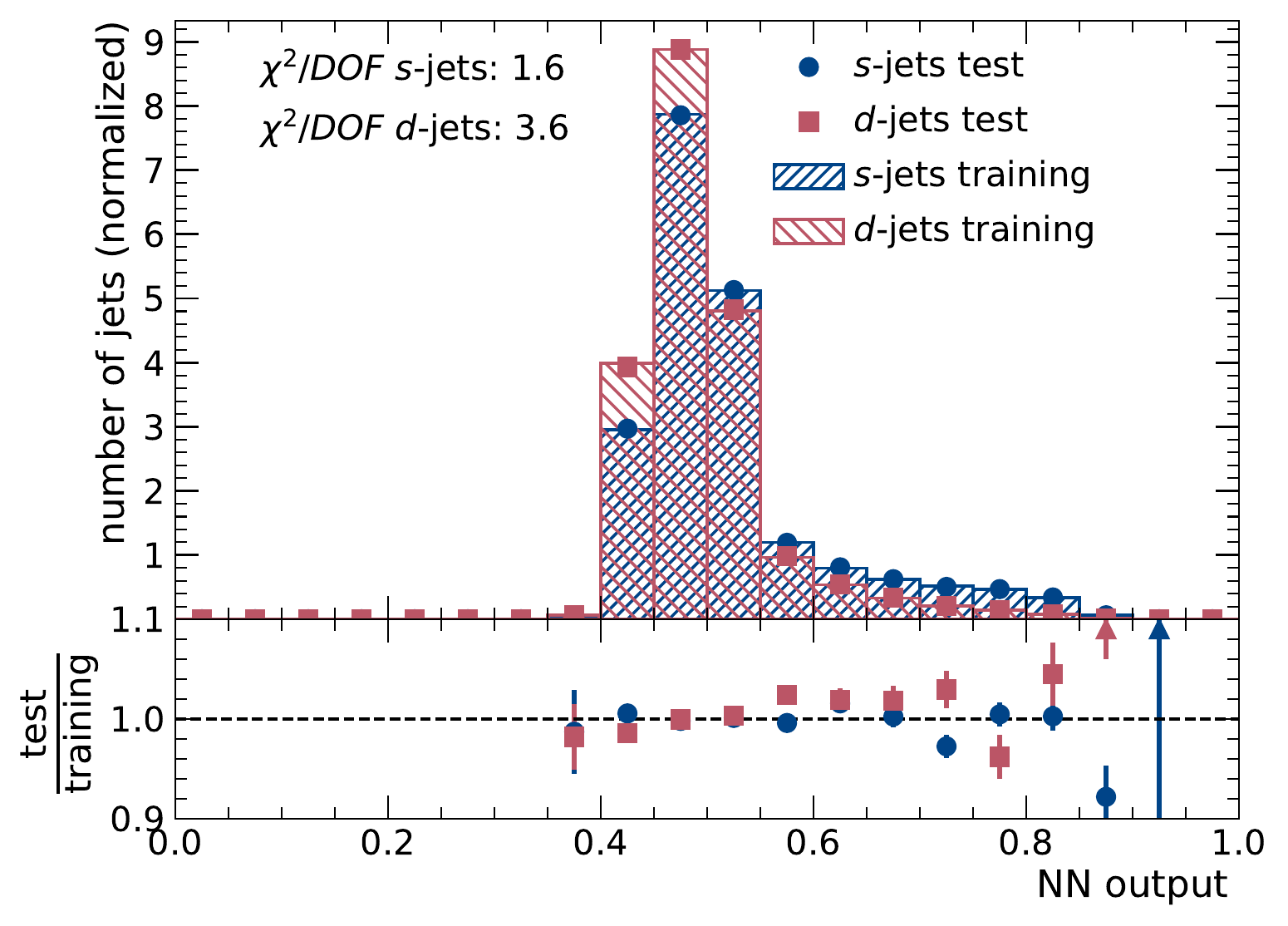}
\caption{Output distribution of the neural network in the tracking detector scenario. The output for the training sample as well as for the independent test sample is compared for $s$-jets and $d$-jets.}
\label{FIG:tracker}
\end{figure}

The output distributions of the neural network in the tracking detector scenario is shown in Figure~\ref{FIG:tracker}. The agreement between the output distribution of the training and the test sample is similar as for the collider detector scenario with a $\chi^2$ value of 1.6 for $s$-jets and 3.6 for $d$-jets per degree of freedom, and only small variations in the ratio of the two distributions.  The separation between $s$-jets and $d$-jets is further reduced with respect to the optimistic collider detector scenario. Most jets respond with a neural network output value around 0.5, which means it is difficult for the neural network to classify these jets with the given information.

\begin{figure}[t]
\centering
\includegraphics[width=.6\textwidth]{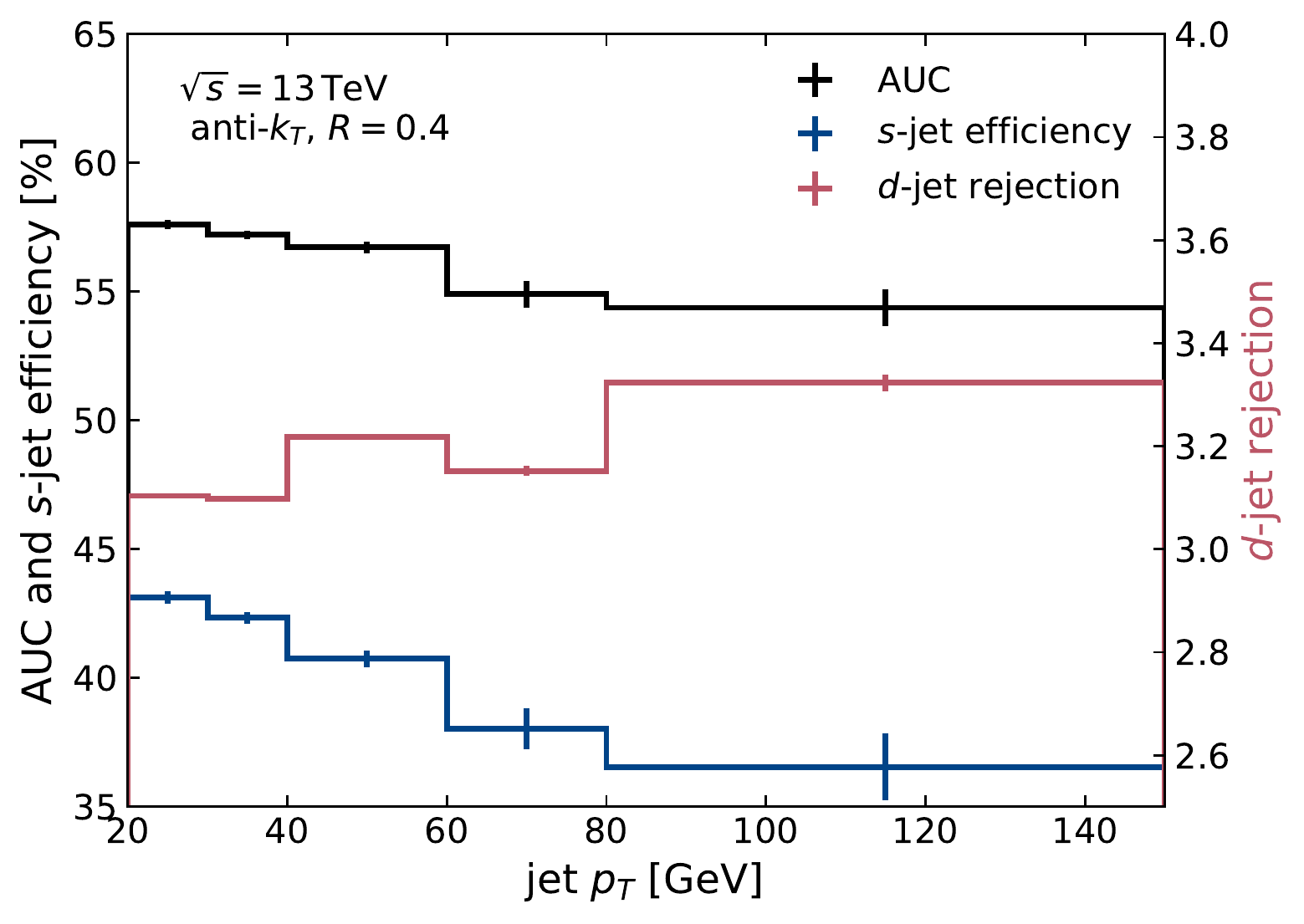}
\caption{AUC, $s$-jet efficiency, and $d$-jet rejection for different jet $p_{\mathrm{T}}$ considering the neural network scenario trained using charged particles only. The efficiency and rejection are calculated in the test sample requiring a neural network output value larger 0.5 for a jet to pass the selection. All uncertainties are statistical uncertainties of the test sample.}
\label{FIG:tracker_pt}
\end{figure}

Figure~\ref{FIG:tracker_pt} illustrates the jet-$p_{\mathrm{T}}$ dependence of the classification. The AUC score decreases slightly for larger $p_{\mathrm{T}}$ as both types of jets become more similar. The figure additionally shows the efficiency for $s$-jets that is decreasing for larger $p_{\mathrm{T}}$ and rejection rate for $d$-jets that is increasing very slightly for larger $p_{\mathrm{T}}$ when a cut value of 0.5 is used for the classification.

\begin{figure}[t]
\centering
\includegraphics[width=.6\textwidth]{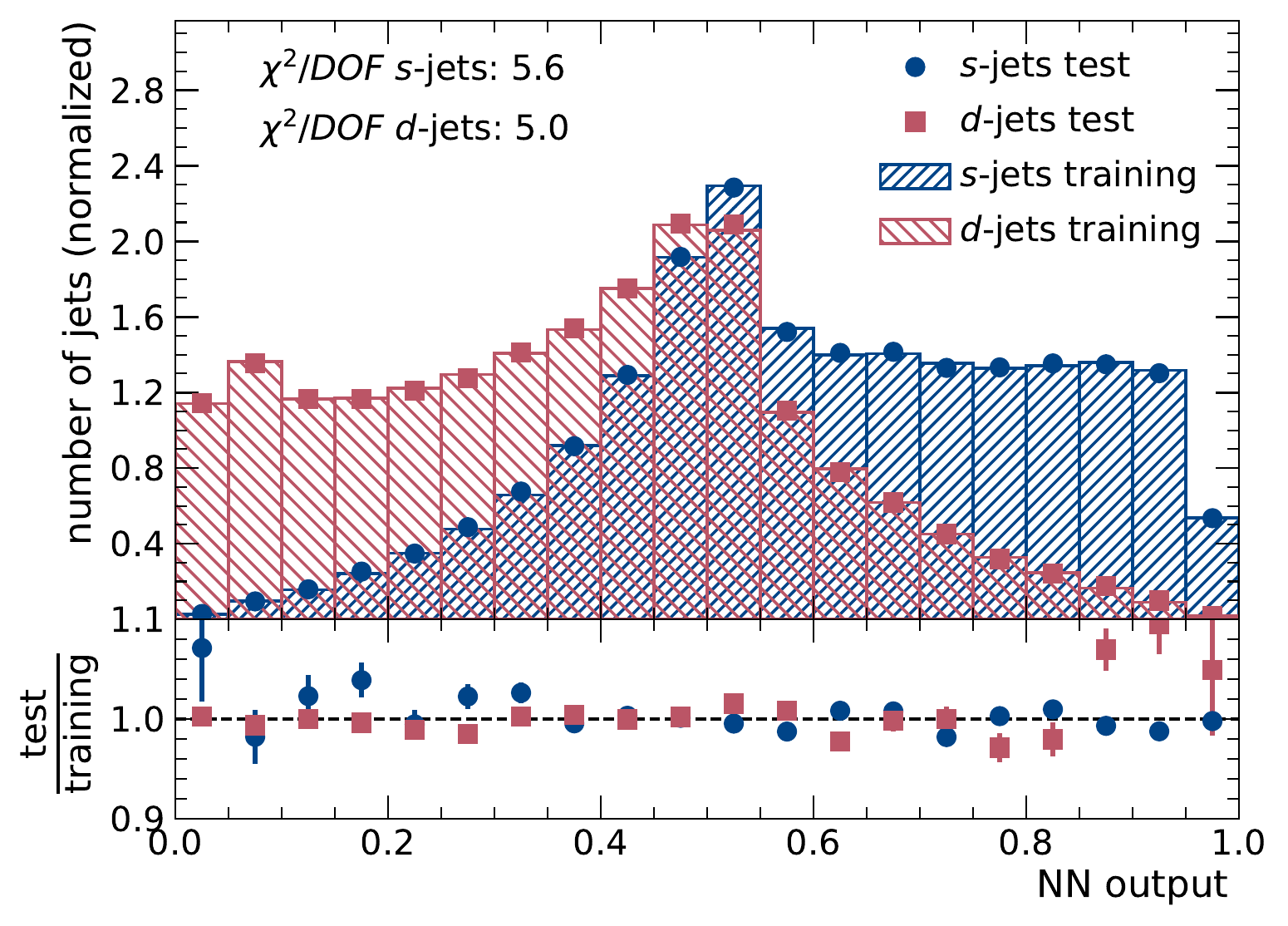}
\caption{Output distribution of the neural network in the Cherenkov detector scenario. The output for the training sample as well as for the independent test sample is compared for $s$-jets and $d$-jets.}
\label{FIG:cherenkov}
\end{figure}

While they are not used in current collider detectors, Cherenkov detectors have been previously exploited for the identification of $s$-jets. The Cherenkov detector scenario evaluates the impact an identification of electrically charged particles such as $K^\pm$ and $\pi^\pm$ with a Cherenkov detector might have, if provided with a sufficient resolution for high-momentum particles. Figure~\ref{FIG:cherenkov} shows the output distribution of the neural network trained with the particles' mass as additional input. The output distribution shows little indication of overtraining in the ratio of the training and the test sample. It is significantly broader than the one for the tracking detector, covering the entire range from 0.0 to 1.0, which indicates that the classification of the two jet types is better.  

\begin{figure}[t]
\centering
\includegraphics[width=.6\textwidth]{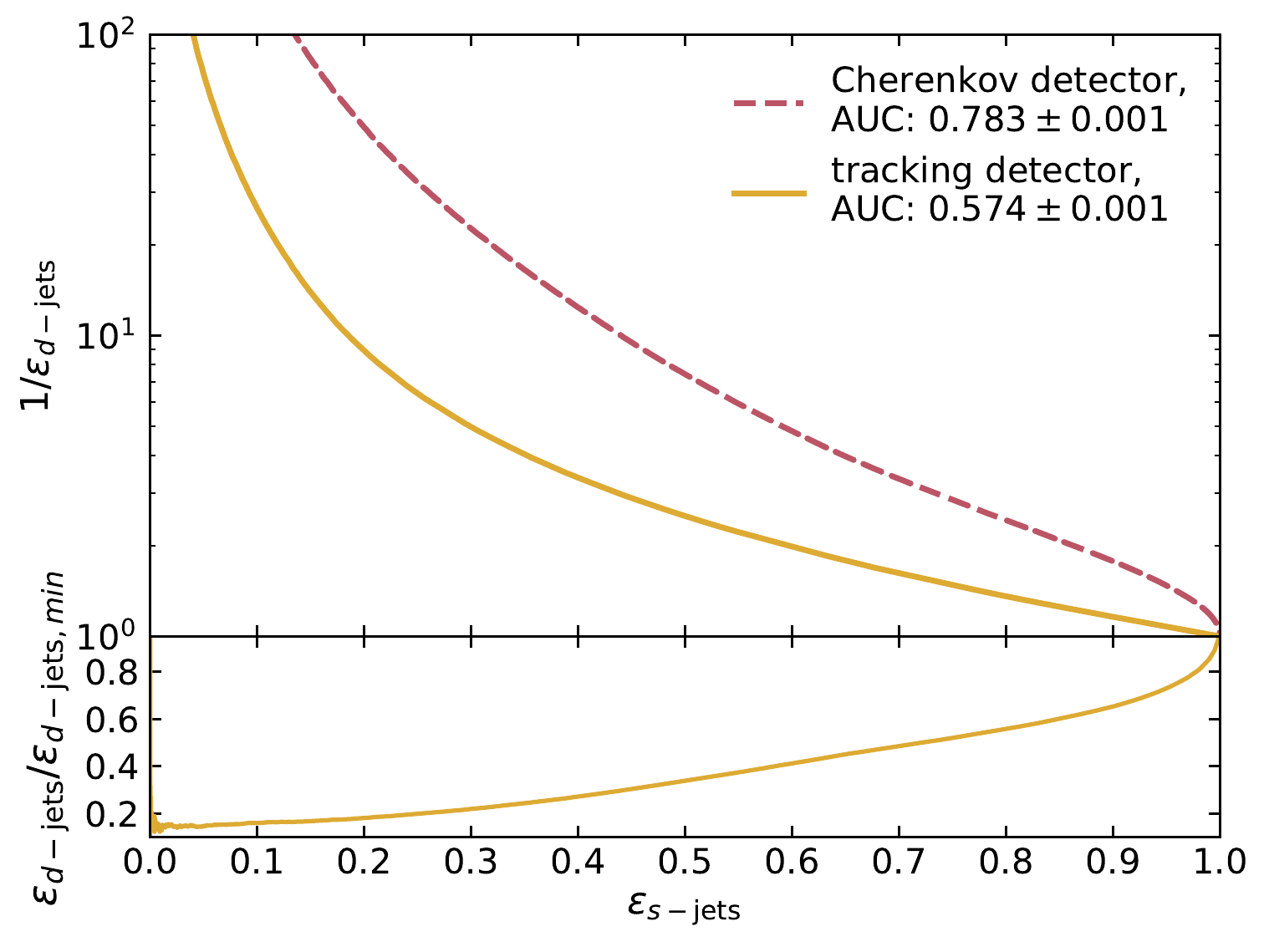} 
\caption{ROC curves illustrating the classification power of neural networks of tracking detector and the Cherenkov detector. The signal is composed of $s$-jets, and the background is composed of $d$-jets. The ratio beneath the ROC curves shows the efficiency for $d$-jets in one scenario ($\varepsilon_{d\text{-jets}}$) divided by the efficiency for $d$-jets in the scenario with the best separation power shown in the ROC curve  ($\varepsilon_{d\text{-jets,min}}$).
The efficiencies are evaluated on the test sample and the uncertainty in the area under the curve is the statistical uncertainty associated with that sample.}
\label{FIG:ROCtrack}
\end{figure}

The ROC curves of both tracking scenarios in Figure~\ref{FIG:ROCtrack} quantify this increase in performance. The ROC curve of the Cherenkov detector shows larger rejection rates than the ROC curve of the tracking detector for all efficiencies. While the tracking scenario has an AUC of $0.574\pm0.001$, the Cherenkov detector has an AUC of $0.783\pm0.001$. Particle identification therefore offers a large potential to improve an $s$-tagging algorithm.

Table~\ref{TAB:Tracker} compares the $d$-jet rejection rates for benchmark $s$-jet efficiencies. In Ref.~\cite{Erdmann:2019blf} the $s$-tagging was also performed based on tracks, but considering in addition a simplified detector simulation with Delphes~\cite{deFavereau:2013fsa}. The rejection rates achieved in this paper are slightly larger, which can be attributed to the finite tracking efficiencies --- which vary between 60\% and 95\% depending on the track's $p_{\text{T}}$ and $\eta$ --- as well as resolution effects.

\begin{table}[h]
\centering
\begin{tabular}{c|c|c}
& $s$-jet efficiency & $d$-jet rejection \\ 
\toprule
tracker scenario & 30\% & 4.9  \\ 
 & 70\% & 1.6\\ 
\midrule
Ref.~\cite{Erdmann:2019blf} & 30\% & 4.5 \\ 
 & 70\% & 1.56 \\ 
\end{tabular} 
\caption{Comparison of $s$-jet efficiencies and $d$-jet rejection rates of neural networks trained in the perfect tracker scenario and using a realistic tracker simulation, cf.~Ref.~\cite{Erdmann:2019blf}.\label{TAB:Tracker}}
\end{table}

\subsection{Calorimeters}

\begin{figure}[t]
\centering
\includegraphics[width=.6\textwidth]{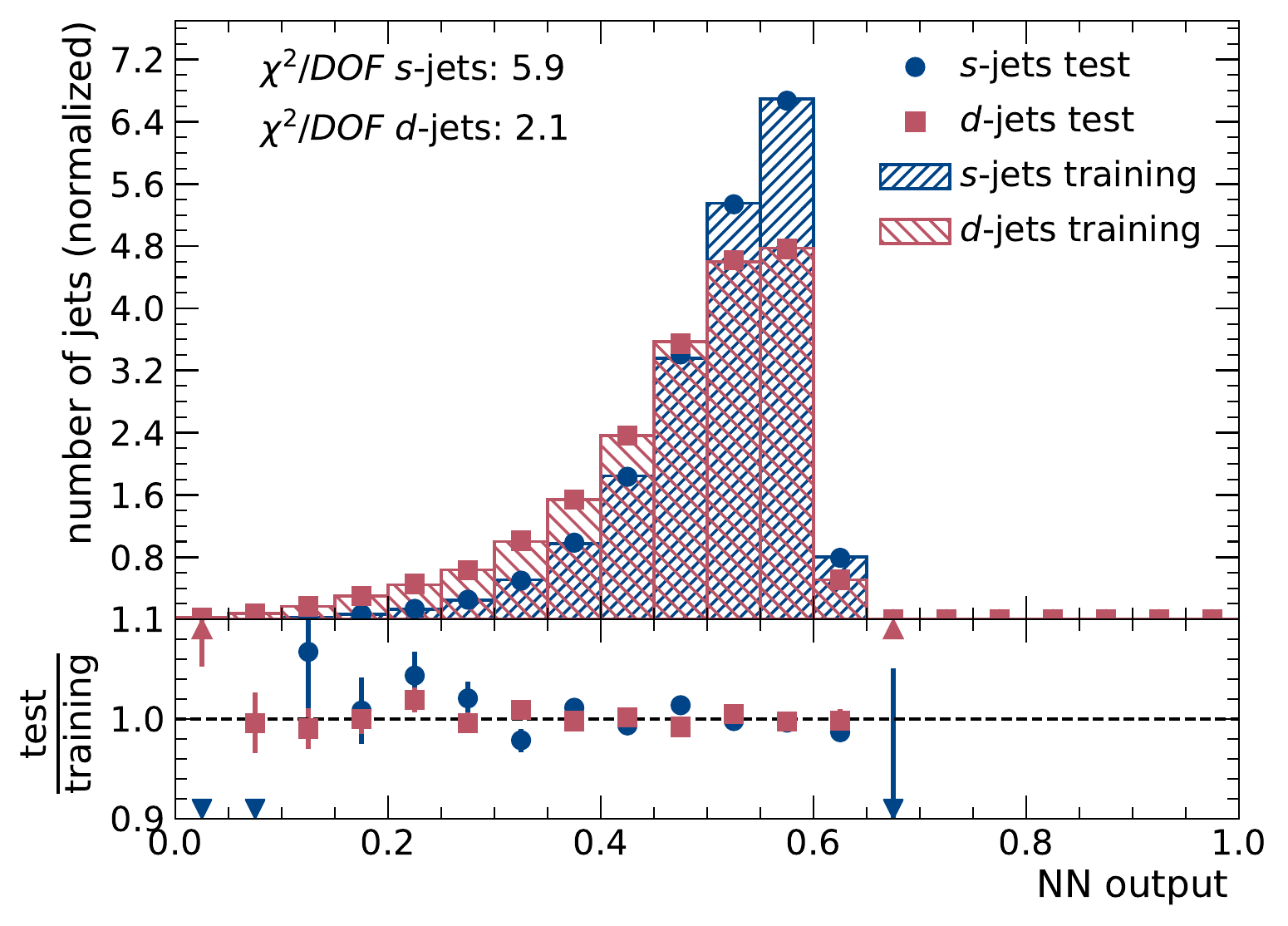}
\caption{Output distribution of the neural network in the calorimeter scenario with separate ECAL and HCAL components. The output for the training sample as well as for the independent test sample is compared for $s$-jets and $d$-jets.}
\label{FIG:calo}
\end{figure}

Figure~\ref{FIG:calo} shows the output distribution of the neural network in the calorimeter scenarios with separation into ECAL and HCAL components. There is no indication of strong overtraining by comparing the response of the test and training set. The separation between $s$-jets and $d$-jets is strongly degraded compared to the previously discussed scenarios, and both jet types respond mostly with an output value around 0.5, which indicates that it is difficult for the neural network to distinguish the two classes.

\begin{figure}[t]
\centering
\includegraphics[width=.6\textwidth]{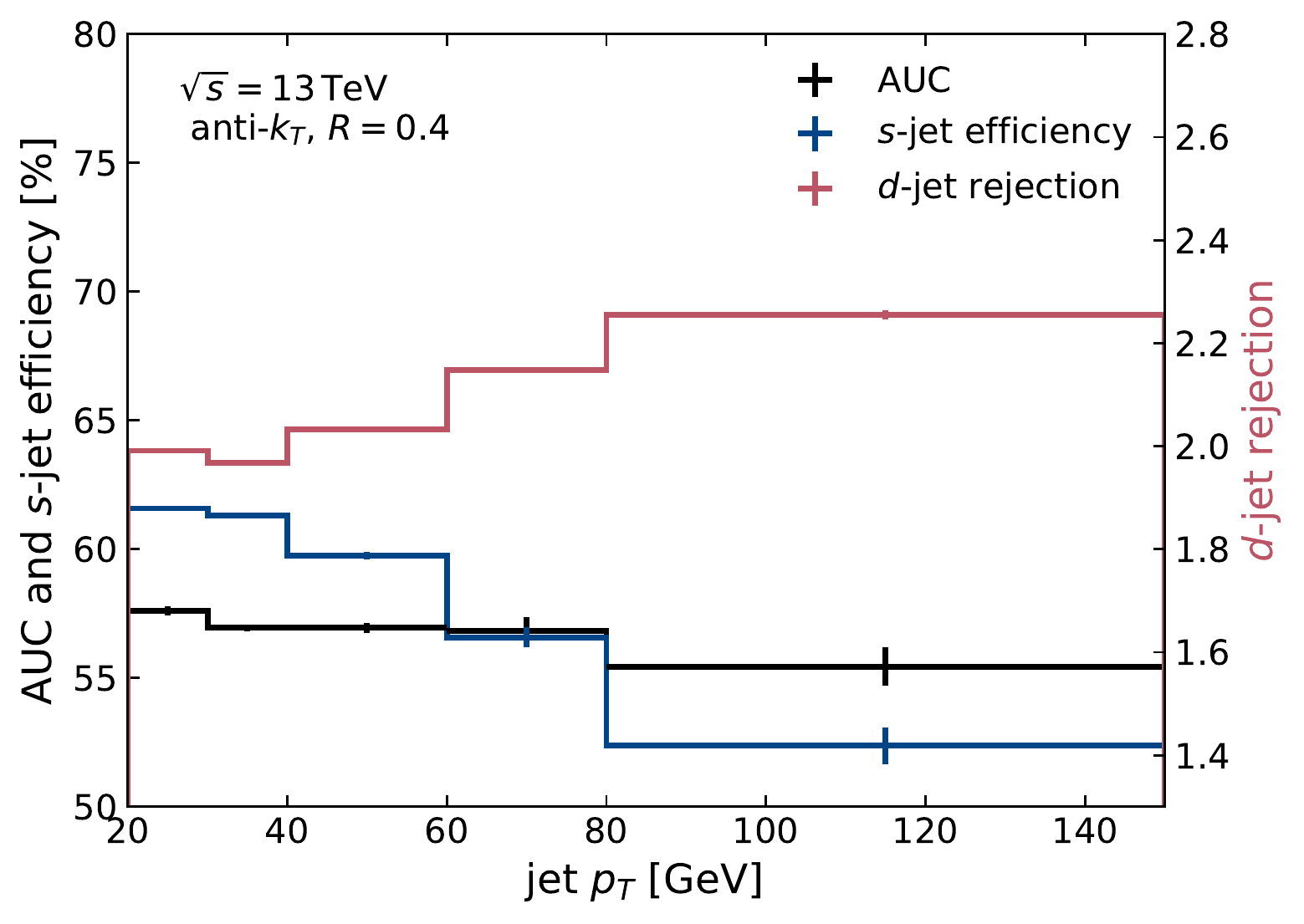}
\caption{AUC, $s$-jet efficiency, and $d$-jet rejection for different jet $p_{\mathrm{T}}$ considering the neural network scenario of the calorimeter divided into an ECAL and HCAL. The efficiency and rejection is calculated in the test sample requiring a neural network output value larger 0.5 for a jet to pass the selection. All uncertainties are statistical uncertainties of the test sample.}
\label{FIG:calo_pt}
\end{figure}

Figure~\ref{FIG:calo_pt} illustrates the jet-$p_{\mathrm{T}}$ dependence of the classification provided by the calorimeter scenario with the division into ECAL and HCAL. The AUC score decreases slightly for larger $p_{\mathrm{T}}$ as both types of jets become more similar. The $s$-jet efficiency when classifying jets with a neural network score larger 0.5 as $s$-jets decreases for larger $p_{\mathrm{T}}$, while the $d$-jet rejection increases slightly. 

\begin{figure}[t]
\centering
\includegraphics[width=.6\textwidth]{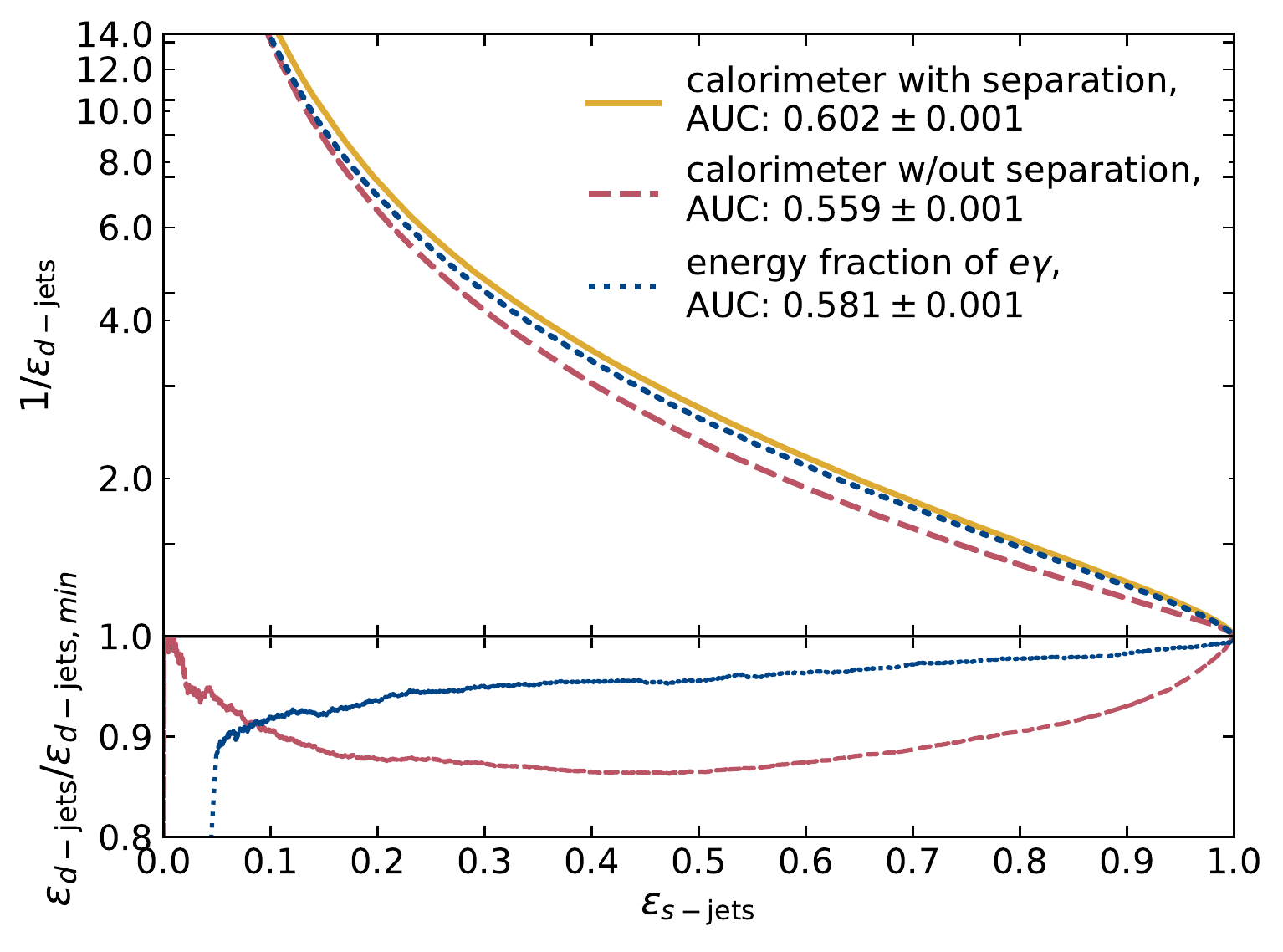}
\caption{ROC curves illustrating the classification power of neural networks in the calorimeter scenarios with and without separation into ECAL and HCAL components, as well as the ROC curve of the fraction of energy deposited by electrons and photons. The signal is composed of $s$-jets, and the background is composed of $d$-jets. 
The ratio beneath the ROC curves shows the efficiency for $d$-jets in one scenario ($\varepsilon_{d\text{-jets}}$) divided by the efficiency for $d$-jets in the scenario with the best separation power shown in the ROC curve  ($\varepsilon_{d\text{-jets,min}}$).
The efficiencies are evaluated on the test sample and the uncertainty in the area under the curve is the statistical uncertainty associated with that sample.}
\label{FIG:ROCcalo}
\end{figure}

Figure~\ref{FIG:ROCcalo} compares the ROC curves of both calorimeter scenarios with a ROC curve obtained only from the distribution of the fraction of energy deposited by electrons and photons, as shown in Figure~\ref{FIG:EMFrac}. The classification power of the network trained with the scenario without a calorimeter separation is significantly smaller with an AUC value of $0.559\pm0.001$ than the one trained with ECAL and HCAL components, which results in an AUC value of $0.602\pm0.001$. This indicates that the separation power is mostly driven by the share of energies between the ECAL and HCAL components instead of the spatial pattern of the energy depositions. This hypothesis is supported by the ROC curve that is constructed only from the energy fraction of electrons and photons, which results in an AUC value of $0.581\pm0.001$, which is almost as good as the one of the ideal calorimeter scenario with a separation into ECAL and HCAL.

\subsection{Strange-hadron decays to two charged particles}

\begin{figure}[t]
\centering
\includegraphics[width=.6\textwidth]{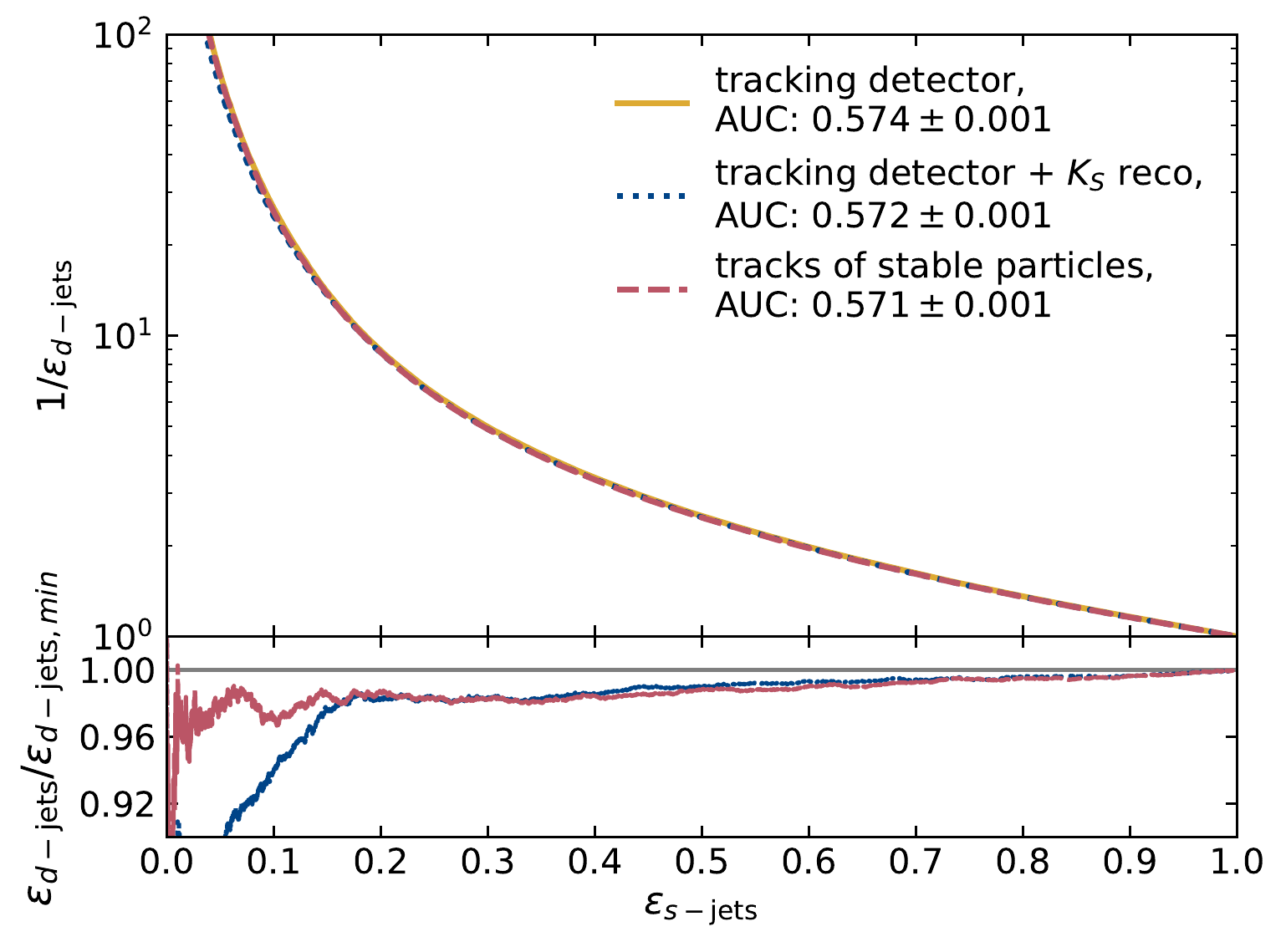}
\caption{ROC curves illustrating the classification power of the tracking detector scenario in comparison to scenarios where tracks of opposite-sign charge from the same origin are combined into a neural particle used as input. Additionally, the same tracker scenario only using stable particles as input is shown.
The ratio beneath the ROC curves shows the efficiency for $d$-jets in one scenario ($\varepsilon_{d\text{-jets}}$) divided by the efficiency for $d$-jets in the scenario with the best separation power shown in the ROC curve  ($\varepsilon_{d\text{-jets,min}}$).
The efficiencies are evaluated on the validation test sample and the uncertainty in the area under the curve is the statistical uncertainty associated with that sample.}
\label{FIG:ROCnpp}
\end{figure}

Figure~\ref{FIG:ROCnpp} compares ROC curves illustrating the classification performances of the neural network trained with the input of  the tracker scenario with networks trained with an altered input. In one scenario, particles decays such as $K_S\rightarrow\pi^+\pi^-$ are reconstructed by replacing two particles from the same origin which have opposite-sign electric charge by a neutral particle whose kinematic properties -- including their mass -- are calculated by adding the four-vectors of the charged particles. Through this scenario, it is possible to determine if the reconstruction of such decays aids the classification of $s$-jets. 

Figure~\ref{FIG:ROCnpp} compares the ROC curves of the two scenarios. The performance of the $s$-tagger with previous reconstruction of $K_S\rightarrow\pi^+\pi^-$ decays is slightly better as it aids the neural network to identify these decays.

\FloatBarrier

\section{Conclusions}

Achieving good performance for $s$-tagging, i.e.~the separation of $s$-jets from other light jets, is challenging. This is in part due to similarities in the fragmentation of $s$- and $d$- or $u$-jets and in part due to experimental challenges. In order to separate these two effects, the performance of $s$-tagging based on idealised detector scenarios was studied. Long short-term memory recurrent neural networks were used as a complex model designed to capture the differences in the jet constituents in the different scenarios.

\begin{figure}[t]
\centering
\includegraphics[width=.6\textwidth]{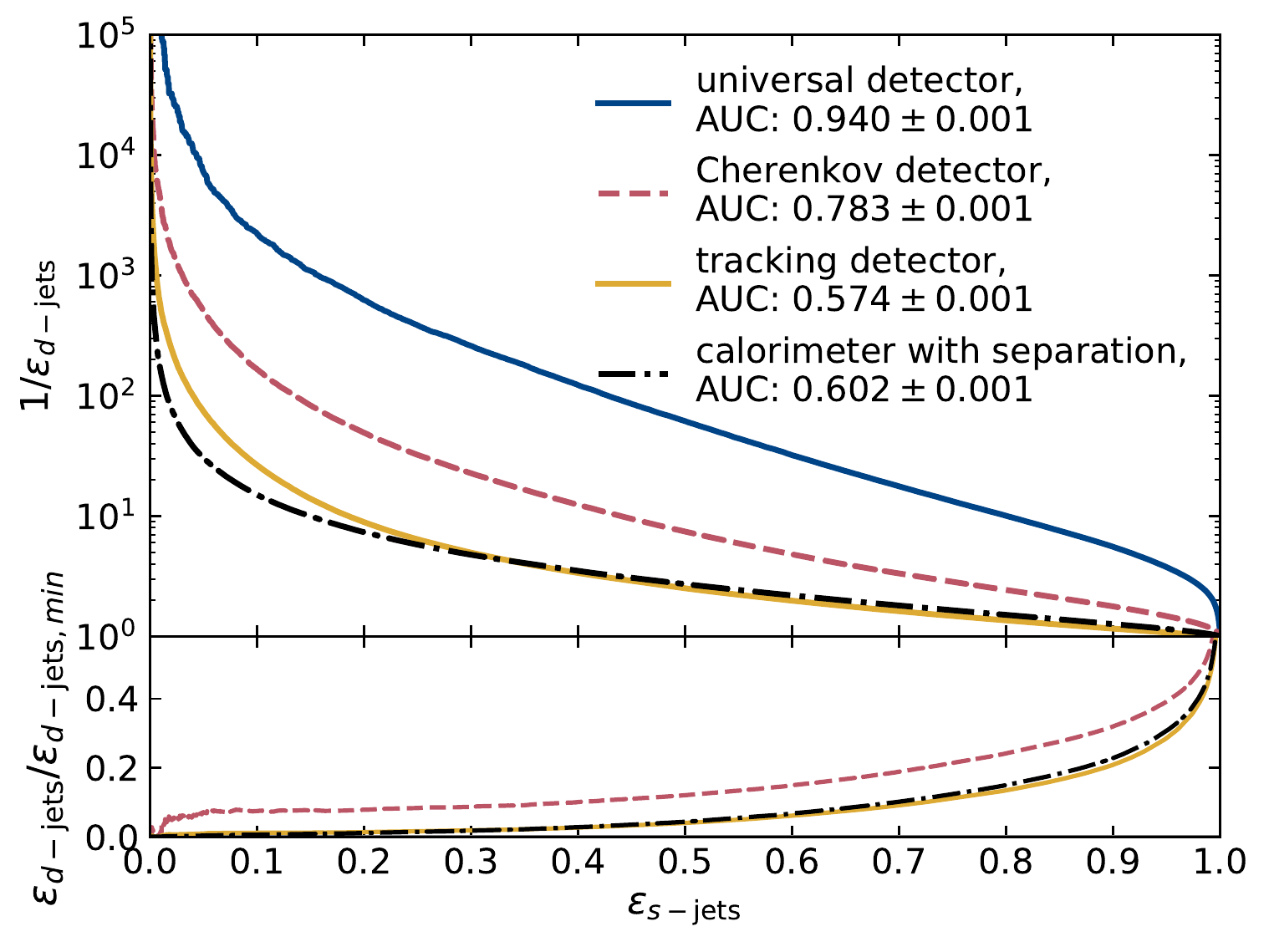}
\caption{ROC curves illustrating the classification power of the neural networks in the universal detector, tracking detector, Cherenkov  detector, and ECAL/HCAL-separated calorimeter scenarios. The signal is composed of $s$-jets, and the background is composed of $d$-jets. 
The ratio beneath the ROC curves shows the efficiency for $d$-jets in one scenario ($\varepsilon_{d\text{-jets}}$) divided by the efficiency for $d$-jets in the scenario with the best separation power shown in the ROC curve  ($\varepsilon_{d\text{-jets,min}}$).
The efficiencies are evaluated on the test sample and the uncertainty in the area under the curve is the statistical uncertainty associated with that sample.}
\label{FIG:ROCcomp}
\end{figure}

An overview of the performance of four main scenarios is shown in Figure~\ref{FIG:ROCcomp}. Assuming an ideal detector that can perfectly measure all jet constituents (``universal collider detector''), $s$- and $d$-jets can be separated well. This means that the fragmentation of $s$- and $d$-jets shows promising differences that may be explored in an $s$-tagging algorithm, but that the maximum achievable performance of an $s$-tagger is by far not as good as for example achieved for $b$-tagging algorithms~\cite{Aad:2019aic,Sirunyan:2017ezt}. 
The comparison also shows that the information provided by high-granularity calorimeters can contribute similarly to $s$-tagging as information from efficient tracking detectors. 
It was shown that the main discriminatory power available in tracks does not stem from the reconstruction of the decays of neutral strange-hadrons, as they represent only a limited fraction of the leading jet constituents in $s$-jets.
Providing additional particle identification capability to the tracking scenario increases the separation of the $s$-tagger significantly, as the main difference between $s$- and $d$-jets is their particle content.

The optimistic collider detector scenario gives an estimate of the maximal performance of an $s$-tagger when combining both, tracking detectors and calorimeters.

Realistic detectors, however, are subject to a reduced fiducial acceptance, material effects, reconstruction inefficiencies, and resolutions for the different input features that are used for the training of the $s$-tagger.  
In particular, track reconstruction is subject to important inefficiencies ---  especially at low track $p_{\text{T}}$ or in dense environments   --- which further reduce the efficacy of a separation based on tracks.
Furthermore, calorimeter measurements are subject to a limited granularity and to significant resolution effects in the energy measurement. Discriminatory power was found both in the  geometric structure of the energy depositions and in the fraction that is carried by electrons and photons. While this fraction is subject to significant fluctuations in the development of hadronic showers (in particular the fraction of high-energy $\pi^0$ mesons, which result in electromagnetic subshowers) and thus has limited resolution, it is a much simpler variable than the complex combination of particle momenta in the LSTM neural network.

While the maximum performance of an $s$-tagging algorithm is limited, its development allows for new opportunities in data analysis at colliders. The results of the ideal detector scenarios underline the importance of following a many-fold approach to $s$-tagging, based on tracking~\cite{Erdmann:2019blf}, calorimeter information~\cite{Nakai:2020kuu}, and -- if available -- particle identification~\cite{Kalelkar:2000ig,Abreu:1999cj}. In order to gauge the performance of $s$-taggers in realistic scenarios, the performance achieved in the scenario with the universal collider detector is proposed as a benchmark, which is solely based on the difference in $s$- and $d$-quark fragmentation.

\acknowledgments

These studies have been financially supported by the Studienstiftung des deutschen Volkes. Special thanks to Nils Julius Abicht for generating the simulated samples used in this study and to Tomas Dado for his comments on the manuscript.

\bibliographystyle{JHEP}
\bibliography{s-tagging}

\end{document}